\documentclass[aps,prd,preprint,superscriptaddress]{revtex4}
\usepackage{ulem}
\usepackage{xcolor}
\usepackage[colorlinks=true]{hyperref}
\usepackage{soul}
\usepackage{booktabs}

\usepackage{amsmath,amssymb,color,epsfig}
\allowdisplaybreaks[4]

\newcommand{\be}{\begin{equation}}
\newcommand{\ee}{\end{equation}}
\newcommand{\bea}{\setlength\arraycolsep{2pt} \begin{eqnarray}}
\newcommand{\eea}{\end{eqnarray}}
\newcommand{\nn}{\nonumber}

\begin{document}

\hypersetup{
    linkcolor=blue,
    citecolor=red,
    urlcolor=magenta
}

% Use the \preprint command to place your local institutional report
% number in the upper righthand corner of the title page in preprint mode.
% Multiple \preprint commands are allowed.
% Use the 'preprintnumbers' class option to override journal defaults
% to display numbers if necessary
%\preprint{}

%Title of paper
\title{Radial oscillations of neutron stars in Starobinsky gravity and its Gauss-Bonnet extension}

% repeat the \author .. \affiliation  etc. as needed
% \email, \thanks, \homepage, \altaffiliation all apply to the current
% author. Explanatory text should go in the []'s, actual e-mail
% address or url should go in the {}'s for \email and \homepage.
% Please use the appropriate macro foreach each type of information

% \affiliation command applies to all authors since the last
% \affiliation command. The \affiliation command should follow the
% other information
% \affiliation can be followed by \email, \homepage, \thanks as well.

\author{Ziyi Li}
\affiliation{Department of Physics, Synergetic Innovation Center for Quantum Effect and Applications, and Institute of Interdisciplinary Studies, Hunan Normal University, Changsha, 410081, China}

\author{Zhong-Xi Yu}
\affiliation{School of Electronic Engineering, Huainan Normal University, Huainan, 232038, China}

\author{Zhe Luo}
\affiliation{Department of Physics, Synergetic Innovation Center for Quantum Effect and Applications, and Institute of Interdisciplinary Studies, Hunan Normal University, Changsha, 410081, China}

\author{Shoulong Li}
\email[Corresponding author: ]{shoulongli@hunnu.edu.cn}
\affiliation{Department of Physics, Synergetic Innovation Center for Quantum Effect and Applications, and Institute of Interdisciplinary Studies, Hunan Normal University, Changsha, 410081, China}

\author{Hongwei Yu}
\email[]{hwyu@hunnu.edu.cn}
\affiliation{Department of Physics, Synergetic Innovation Center for Quantum Effect and Applications, and Institute of Interdisciplinary Studies, Hunan Normal University, Changsha, 410081, China}

%Collaboration name if desired (requires use of superscriptaddress
%option in \documentclass). \noaffiliation is required (may also be
%used with the \author command).
%\collaboration can be followed by \email, \homepage, \thanks as well.
%\collaboration{}
%\noaffiliation

\date{\today}

\begin{abstract}

Starobinsky gravity, as one of the simplest and best-behaved higher-curvature gravity theories, has been extensively studied in the context of neutron stars over the past few decades. In this work, we investigate the adiabatic radial oscillation stability of neutron stars within the framework of Starobinsky gravity. We find that gravitational modifications can significantly impact stellar stability. Specifically, the higher-derivative nature of the theory causes the exterior spacetime to dynamically respond to fluid oscillations, in contrast to general relativity where Birkhoff's theorem ensures a static exterior.
For stellar models with low central densities, the fundamental frequency becomes nearly independent of the central density when the coupling constant is large. For stellar models with high central densities, the transition from stability to instability still approximately occurs near the maximum-mass configuration, similar to the case in general relativity. Our main analysis is conducted in the Jordan frame of the scalar-tensor gravity equivalent to Starobinsky gravity, and we explicitly verify consistency with results obtained in the Einstein frame. We further extend our study to a class of Gauss-Bonnet extensions of Starobinsky gravity.

\end{abstract}
% insert suggested keywords - APS authors don't need to do this
%\keywords{}

%\maketitle must follow title, authors, abstract, and keywords
\maketitle

\section{Introduction}

Higher-curvature gravity theories~\cite{Padmanabhan:2013xyr, Belenchia:2016bvb, Shankaranarayanan:2022wbx, Sotiriou:2008rp, DeFelice:2010aj, Nojiri:2017ncd}, which extend general relativity (GR) by incorporating curvature invariants, have attracted sustained attention over the past few decades. One primary motivation is their potential renormalizability in four dimensions~\cite{Stelle:1976gc, Stelle:1977ry}, along with their natural connection to high-curvature corrections predicted by string theory and effective field theory~\cite{Callan:1985ia, Donoghue:1994dn}. However, the inclusion of higher-curvature terms may introduce additional massive modes, such as massive scalar and massive spin-2 modes, into the linear spectrum of the maximally symmetric vacuum outside gravitational sources, in addition to the usual massless graviton. Given the ghost-like nature of the massive spin-2 mode, which leads to unphysical instabilities, most viable models are constructed to include only a massive scalar or to avoid any extra degrees of freedom entirely.

Among these gravitational models, Starobinsky gravity~\cite{Starobinsky:1980te}, also known as curvature-squared ($\mathcal{R}^2$) gravity, is one of the simplest and best studied. It provides a self-consistent inflationary mechanism without invoking an ad hoc inflaton field. Beyond its cosmological success, Starobinsky gravity has also been extensively explored in the context of neutron stars~\cite{Cooney:2009rr, Babichev:2009fi, Doneva:2015hsa, Staykov:2015cfa, Ganguly:2013taa, Jaime:2010kn, Arapoglu:2010rz, Orellana:2013gn, Astashenok:2013vza, Numajiri:2023uif, Yazadjiev:2018xxk, Staykov:2015kwa, Staykov:2014mwa, Capozziello:2015yza, Yazadjiev:2014cza, AparicioResco:2016xcm,Fang:2017xyy,Astashenok:2017dpo, Kobayashi:2008tq, Upadhye:2009kt, Babichev:2009td, Sbisa:2019mae, Pretel:2020rqx, Feola:2019zqg,Bonanno:2021zoy,Liu:2024wvw}. For a comprehensive overview of neutron stars in Starobinsky gravity, see the reviews~\cite{Berti:2015itd,Barack:2018yly,Olmo:2019flu}, and the references therein. In particular, the macroscopic properties of neutron stars, such as their mass, radius, moment of inertia, tidal Love number, and quadrupole moment, can deviate significantly from their counterparts in GR. 
Beyond modifying stellar interior structure, Starobinsky gravity also alters the exterior vacuum geometry near the surface of neutron stars. Because the theory introduces a massive scalar mode on a maximally symmetric vacuum background, static and spherically symmetric neutron stars can support scalar hair, leading to departures from the Schwarzschild spacetime of GR. This naturally raises the question of how such modifications impact the dynamical behavior of neutron stars~\cite{Kokkotas:1999bd}, in particular their radial stability under small perturbations.

Previous studies~\cite{Pretel:2020rqx} have approached this question by interpreting the curvature-squared term as an effective curvature fluid. This leads to the concept of a ``gravitational sphere''~\cite{Astashenok:2017dpo}, defined as a region with nonvanishing curvature near the stellar surface, in contrast to the vacuum exterior of a Schwarzschild black hole. Although this interpretation may offer insights into equilibrium configurations in Starobinsky gravity, it does not readily generalize to broader classes of modified gravity theories. For instance, in Gauss-Bonnet extended Starobinsky gravity, both neutron stars~\cite{Liu:2024wvw} and black holes~\cite{Liu:2020yqa} can support massive scalar hair, with their exterior spacetime exhibiting nonzero curvature, while the Schwarzschild black hole remains a special case with vanishing exterior curvature. (Similar types of hairy solutions also arise in general quadratic gravity, where massive spin-2 hair can be supported~\cite{Lu:2015cqa,Chen:2024hsh}.) Furthermore, in ghost-free cubic quasi-topological gravity~\cite{Li:2017ncu, Hennigar:2017ego}, even neutron stars without scalar hair can exhibit nonzero vacuum curvature near their surfaces~\cite{Li:2023vbo}, although the Schwarzschild solution still exists as a special case with vanishing curvature outside the horizon. These examples indicate that the notion of a gravitational sphere is not a generic feature of higher-curvature theories. A more consistent and universal explanation for the nonvanishing exterior curvature lies in the breakdown of Birkhoff's theorem~\cite{Liu:2024wvw}, which accounts for non-Schwarzschild geometries in static, spherically symmetric spacetimes.

Motivated by these observations, we note that extending the effective curvature fluid interpretation to studies of dynamical behavior under radial perturbations may introduce conceptual and methodological ambiguities. In particular, under this interpretation~\cite{Pretel:2020rqx}, the location where the Lagrangian perturbation of pressure vanishes may differ from the stellar surface as defined in the conventional GR framework, potentially affecting the formulation of boundary conditions and stability criteria. To avoid such issues, we revisit the radial stability of neutron stars in Starobinsky gravity using the standard formalism in GR~\cite{Misner:1973prb, Chandrasekhar:1964zza, Kokkotas:2000up}. We further extend our analysis to include the Gauss-Bonnet extension of Starobinsky gravity~\cite{Liu:2020yqa},  thereby exploring a broader class of higher-curvature gravities beyond the assumptions of a single model.

The remainder of this paper is structured as follows. In Sec.~\ref{framework}, we begin with a brief review of the gravitational framework, including Starobinsky gravity and its Gauss-Bonnet extension. We then present the corresponding field equations, derive the modified Tolman-Oppenheimer-Volkoff (TOV) equilibrium equations, and formulate the radial perturbation equations along with the associated boundary conditions. In Sec.~\ref{Stability}, we examine the dynamical behavior of the solutions and analyze their stability. Finally, in Sec.~\ref{Conclusion}, we summarize our main results and provide concluding remarks.

\section{Framework} \label{framework}

In this section, we begin with a brief review of Starobinsky gravity and its Gauss-Bonnet extension, along with the corresponding equations of motion (EOMs). We then derive the equations governing the radial oscillations of neutron stars within these gravitational frameworks.

\subsection{Starobinsky gravity and its Gauss-Bonnet extension} 

For any gravitational theory, the complete action $S$ can be expressed as
\be
S = \frac{c^4}{16 \pi G} \int{ d^4 x \sqrt{-g} L} +S_\textup{m}(g_{\mu\nu}, \chi)  \,, \label{action}
\ee
where $g$, $L$, and $S_\textup{m}$ represent determinant of metric $g_{\mu\nu}$, the gravitational Lagrangian density, and the action for the matter field $\chi$, respectively. The symbols $c$ and $G$, representing the speed of light and the gravitational constant, are set to unity ($c=G=1$) in the remainder of the work, as we adopt geometric units. The specific expression for $L$ in Starobinsky gravity and its Gauss-Bonnet extension takes the following form~\cite{Liu:2020yqa}: 
\be
L = {\cal R} + \frac{\alpha {\cal R}^2}{1 -2 \alpha \beta L_\textup{GB}} \,, \quad \textup{with}  \quad L_\textup{GB} = {\cal R}^2 - 4 {\cal R}_{\mu\nu}{\cal R}^{\mu\nu} +  {\cal R}_{\mu\nu\rho\sigma}{\cal R}^{\mu\nu\rho\sigma} \,, \label{GBES}
\ee
where ${\cal R}$, ${\cal R}_{\mu\nu}$, and ${\cal R}_{\mu\nu\rho\sigma}$ represent Ricci scalar, Ricci tensor, and Riemann tensor, respectively.  Both parameters $\alpha$ and $\beta$ are coupling constants. Setting $\beta = 0$ eliminates the contribution of the Gauss-Bonnet coupling, reducing the theory to the celebrated Starobinsky gravity~\cite{Starobinsky:1980te}. Further setting $\alpha = 0$ returns the theory to standard GR. 
The Lagrangian~(\ref{GBES}) has a specific equivalent form in scalar-tensor gravity, which can be written as
\be
L = {\cal R} +\Phi {\cal R} -\frac{1}{2} \mu^2 \Phi^2 +U(\Phi) L_\textup{GB}\,, \label{edgb}
\ee
with 
\be
U = \frac{1}{2}\beta \Phi^2 \,, \quad \mu^2 = \frac{1}{2\alpha} >0 \,,
\ee
where $\Phi$ represents a massive scalar field, and $\mu$ can be understood as its mass. 
When $\beta = 0$, the Lagrangian~(\ref{edgb}) is typically regarded as the form of the Brans-Dicke theory with a specific parameter $w_\textup{BD}=0$ in Jordan frame. In this sense, the full Lagrangian~(\ref{edgb}) can be understood as the Jordan-frame representation of the scalar-tensor gravity equivalent to Gauss-Bonnet extended Starobinsky gravity. The EOMs $E_{\mu\nu}$ and $E_\Phi$ are given by
\bea
E_{\mu\nu} &\equiv& (\Phi +1) G_{\mu\nu} + g_{\mu\nu} \Box \Phi - \nabla_\mu\nabla_\nu \Phi + \frac{\mu^2}{4} \Phi^2 g_{\mu\nu} + X_{\mu\nu}  = 8\pi T_{\mu\nu} \,, \label{eom1} \\
E_\Phi &\equiv& R -\mu^2 \Phi + \beta \Phi L_\textup{GB} = 0 \,, \label{eom2}
\eea
where $G_{\mu\nu} = {\cal R}_{\mu\nu} - {\cal R} g_{\mu\nu}/2$ is the Einstein tensor, $\nabla_\mu$ is the covariant derivative, $ \Box = \nabla_\mu \nabla^\mu$ is the d'Alembert operator, and $X_{\mu\nu}$ and the energy momentum tensor $T_{\mu\nu} $ are expressed as
\bea
X_{\mu\nu} &=& 8 {\cal R}^\rho{}_{(\mu} \nabla_{\nu)}  \nabla_\rho U - 2 {\cal R} \nabla_\mu \nabla_\nu U  - 4 G_{\mu\nu} \Box U  + 4 {\cal R}^\rho{}_{\mu}{}^\sigma{}_{\nu} \nabla_\rho \nabla_\sigma U - 4 g_{\mu\nu} {\cal R}^{\rho\sigma} \nabla_\rho \nabla_\sigma U \,, \\
T_{\mu\nu} &=& -\frac{2}{\sqrt{-g}} \frac{\partial S_\textup{m}}{\partial g^{\mu\nu}} \,.
\eea
Here, the parentheses denote the symmetrization of the indices enclosed within them.
By solving Eq.~(\ref{eom2}), we obtain 
\be
\Phi = \frac{2\alpha {\cal R}}{1- 2\alpha \beta L_\textup{GB}} \,. \label{phicurv}
\ee
Note that the Ricci scalar $\cal R$ is linearly related to scalar field $\Phi$ only in Starobinsky gravity, but not in its Gauss-Bonnet extension. Substituting the above equation into the Lagrangian~(\ref{edgb}), it can be verified that the Lagrangian reduces back to Eq.~(\ref{GBES}). 
In the following discussion, we will use the Lagrangian~(\ref{edgb}) and its EOMs~(\ref{eom1})--(\ref{eom2}) to study the radial oscillations of neutron stars.
Notably, Starobinsky gravity can be equivalently formulated in both the Jordan and Einstein frames as a scalar-tensor gravity. In Appendix~\ref{appendix}, we demonstrate that the radial oscillation behavior of neutron stars remains consistent across these two formulations.

\subsection{Modified Tolman–Oppenheimer–Volkoff equations}

The interior and exterior spacetime of a spherically symmetric star is described by the metric ansatz
\be
ds^2 = - e^{\tilde{\lambda}(t, r)} dt^2 + \tilde{f}(t, r)^{-1} dr^2  +r^2 (d\theta^2 + \sin^2\theta d\varphi^2)  \,, \label{metric}
\ee
in Schwarzschild coordinates $(t, r, \theta, \varphi)$, where $\tilde{\lambda}$, and $\tilde{f}$ are metric functions. Throughout the following discussion, functions denoted with a tilde depend on both the temporal and radial coordinates, $t$ and $r$.
The dynamical scalar field $\Phi$ is represented as 
\be
\Phi = \tilde{\phi}(t, r) \,.\label{scalar}
\ee
Assuming the star is composed of a perfect fluid, the energy-momentum tensor takes the standard form
\be
T^{\mu\nu} = (\tilde{\rho} (t, r) + \tilde{p} (t, r)) u^\mu u^\nu + \tilde{p} (t, r) g^{\mu\nu} \,, \label{tmunu}
\ee
where $\tilde{\rho}$ and $\tilde{p}$ denote the density and pressure, respectively, and the four-velocity of the fluid $u^\mu$ is defined by
\be
u^\mu = \frac{dx^\mu}{d\tau} \,, \label{udef}
\ee
with $\tau$ being the proper time, and satisfies the normalization condition
\be
u^\mu u_\mu = -1 \,. \label{ucond}
\ee
The fluid obeys an equation of state (EOS) of the form
\be
\tilde{p} = P ( \tilde{\rho} ) \,. \label{eos}
\ee

Under radial perturbations that preserve spherical symmetry, a fluid element initially located at radial coordinate $r$ in the equilibrium configuration is displaced to $r + \epsilon \delta r(t, r)$ at time $t$ in the perturbed configuration. Here $\epsilon$ is a bookkeeping parameter, and  $\delta r$ represents the radial displacement. The corresponding perturbed quantities---namely, metric functions $\tilde{\lambda}$ and $\tilde{f}$, scalar field $\tilde{\phi} $,  pressure $\tilde{p}$,  density $\tilde{\rho}$, and four-velocity $\tilde{u}^\mu$---are expanded to first order in $\epsilon$ as
\begin{subequations} \label{functionpert}
\bea
\tilde{\lambda}(t, r) &=& \lambda(r) + \epsilon \delta \lambda (t, r) \,, \label{lambdapert} \\
\tilde{f}(t, r) &=& f(r) + \epsilon \delta f (t, r) \,,  \label{fpert} \\
\tilde{\phi}(t, r) &=& \phi(r) + \epsilon \delta \phi (t, r) \,, \label{phipert}  \\
\tilde{p}(t, r) &=& p(r) + \epsilon \delta p (t, r) \,,  \label{ppert} \\
\tilde{\rho}(t, r) &=& \rho(r) + \epsilon \delta \rho (t, r) \,, \label{rhopert}  \\
\tilde{u}^\mu(t, r) &=& u^\mu(r) + \epsilon \delta u^\mu (t, r) \,. \label{upert}
\eea
\end{subequations}
Here, $\lambda, f, \phi, p, \rho$, and $u^\mu$ represent the physical quantities in the equilibrium configurations, while the symbol with a ``$\delta$'' represent the Eulerian perturbations~\cite{Misner:1973prb}, i.e., changes measured by an observer at fixed spatial coordinates. 

Using Eqs.~(\ref{udef}) and (\ref{upert}), we find
\be
\frac{\tilde{u}^r}{\tilde{u}^t} = \frac{dr/d\tau}{dt/d\tau} = \frac{dr}{dt}= \epsilon \dot{\delta r} \,, \quad \frac{\tilde{u}^\theta}{\tilde{u}^t} =  \frac{\tilde{u}^\varphi}{\tilde{u}^t} = 0 \,,\label{uvelocity}
\ee
where the dot denotes the derivative with respect to $t$. Substituting into Eq.~(\ref{ucond}) and expanding to first order in $\epsilon$ yields the unperturbed and perturbed four-velocity:
\bea
u^\mu &=&  (e^{-\frac{\lambda}{2}}, 0, 0, 0) \,, \label{velocity0} \\
\delta u^\mu &=& (-\frac{e^{-\frac{\lambda}{2}}}{2} \delta \lambda, e^{-\frac{\lambda}{2}} \dot{\delta r}, 0, 0)\,. \label{velocity1}
\eea
Using the EOS~(\ref{eos}) and expanding pressure and density as in Eqs.~(\ref{ppert})--(\ref{rhopert}), we have
\bea
p &=& P(\rho) \,, \label{eos0}\\
\delta p &=& \frac{\partial p}{\partial \rho} \delta \rho \label{eos1}\,.
\eea
Inserting Eqs.~(\ref{velocity0})--(\ref{eos1}) into the expressions for the perturbed quantities in (\ref{functionpert}) and substituting them into the energy-momentum tensor (\ref{tmunu}) gives both the background and perturbed forms of $T^{\mu\nu}$. Substituting the metric~(\ref{metric}), scalar field~(\ref{scalar}), and $T^{\mu\nu}$ into the EOMs~(\ref{eom1}) --(\ref{eom2}), and expanding in orders of $\epsilon$, we obtain the background equilibrium equations (i.e., the modified TOV equations) and the perturbation equations governing radial oscillations.

In the case $\beta = 0$, corresponding to Starobinsky gravity, the ${\cal O} (\epsilon^0)$ components of the field equations for $E^\mu{}_\nu$ and $E_\Phi$ reduce to
\begin{subequations} \label{TOV}
\bea
E^t{}_t &\equiv& \frac{\phi ^2}{8 \alpha }+\frac{(\phi +1) \left(r f'+f-1\right)}{r^2}+\phi ' \left(\frac{f'}{2}+\frac{2 f}{r}\right)+f \phi ''+8 \pi  \rho =0 \,, \label{starobinsky1}\\
E^r{}_r &\equiv& \frac{\phi ^2}{8 \alpha }+\frac{f (\phi +1) \left(r \lambda '+1\right)}{r^2}+\frac{f \phi ' \left(r \lambda '+4\right)}{2 r}-8 \pi  P(\rho)-\frac{\phi }{r^2}-\frac{1}{r^2} =0 \,,\label{starobinsky2}\\
E^\theta{}_\theta &\equiv& \frac{\phi ^2}{8 \alpha }+\phi ' \left(\frac{f'}{2}+\frac{f \left(r \lambda '+2\right)}{2 r}\right)+\frac{(\phi +1) f' \left(r \lambda '+2\right)}{4 r}+\frac{1}{2} f (\phi +1) \lambda '' \nn\\ 
&\quad& +\frac{f (\phi +1) \lambda ' \left(r \lambda '+2\right)}{4 r}+f \phi ''-8 \pi  P(\rho) =0 \,,\label{starobinsky3}\\
E_\Phi &\equiv& -\frac{\phi }{2 \alpha }-\frac{f' \left(r \lambda '+4\right)}{2 r}-f \lambda ''+\frac{4-f \left(r \lambda '+2\right)^2}{2 r^2} =0 \,. \label{starobinskyphi}
\eea
\end{subequations} 
The scalar field equation~(\ref{starobinskyphi}) can be explicitly solved as
\be
\phi = -\frac{\alpha  \left(r f' \left(r \lambda '+4\right)+f \left(2 r^2 \lambda ''+r^2 \lambda '^2+4 r \lambda '+4\right)-4\right)}{r^2} \,. \label{starobinskyphi2}
\ee
By substituting Eq.~(\ref{starobinskyphi2}) into Eqs.~(\ref{starobinsky1})--(\ref{starobinsky3}) and expanding in powers of the small parameter $\alpha$, it is straightforward to verify that the equations reduce to those of GR in the limit $\alpha \to 0$.
For $\beta \ne 0$, which corresponds to the Gauss-Bonnet extended Starobinsky gravity, the explicit expressions become more involved. The modified TOV equations derived from the ${\cal O}(\epsilon^0)$ components of the field equations form a system of nonlinear ordinary differential equations (ODEs):
\bea
\lambda^{\prime\prime} &=& F_1 (r, \lambda^{\prime}, f, \phi, \rho) \,, \nn\\
f^{\prime} &=& F_2 (r, \lambda^{\prime}, f, \phi, \rho) \,, \nn\\
\phi^{\prime} &=& F_3 (r, \lambda^{\prime}, f, \phi, \rho) \,, \nn\\
\rho^{\prime} &=& F_4 (r, \lambda^{\prime},  \rho) \,,  \label{zeroorder}
\eea
where the functions $F_i$ (with $i=1, 2, 3, 4$) encapsulate the lengthy and nonlinear dependencies of the system. Their explicit forms, also presented in Appendix A of Ref.~\cite{Liu:2024wvw}, are included in the supplemental material for completeness. 

In the vacuum regions, both the pressure and density vanish, i.e., $\rho = 0$ and $p = 0$. 
By substituting these conditions into the system of equations~(\ref{zeroorder}), the field equations reduce to a system of four first-order ODEs, which take the form:
\bea
\lambda^{\prime\prime} &=& \tilde{F}_1 (r, \lambda^{\prime}, f, \phi) \,, \nn\\
f^{\prime} &=& \tilde{F}_2 (r, \lambda^{\prime}, f, \phi) \,, \nn\\
\phi^{\prime} &=& \tilde{F}_3 (r, \lambda^{\prime}, f, \phi) \,.  \label{zeroordervac}
\eea

To solve the equilibrium equations~(\ref{zeroorder}) and the corresponding  vacuum equations~(\ref{zeroordervac}), it is necessary to specify appropriate boundary conditions.
Near the center of the star, the behavior of the regular solution can be described by power series expansions of the equilibrium functions $\lambda$, $f$, $\phi$, and $\rho$ as follows:
\bea
\lim_{r \rightarrow 0} \lambda(r) &=& \lambda_0 +\lambda_1 r +\lambda_2 r^2 +{\cal O}(r^3) \,, \nn \\
\lim_{r \rightarrow 0} f (r) &=& f_0 + f_1 r+ f_2 r^2 +{\cal O}(r^3)\,, \nn \\ 
\lim_{r \rightarrow 0} \phi(r) &=& \phi_0 + \phi_1 r + \phi_2 r^2 +{\cal O}(r^3)\,, \nn \\
\lim_{r \rightarrow 0} \rho(r) &=& \rho_0 + \rho_1 r +{\cal O}(r^2)\,.  \label{centercond1}
\eea
The independent parameters in these expansions include the central density $\rho_0$ and the central values $\lambda_0$ and $\phi_0$. The nonzero coefficients $\lambda_2$, $f_0$, $f_2$, and $\phi_2$ are determined by the field equations and can be expressed in terms of these free parameters. Their explicit forms are provided in Appendix A of Ref.~\cite{Liu:2024wvw}. In the special case $\beta = 0$, corresponding to Starobinsky gravity,  the coefficients simplify to the following forms:
\bea
\lambda_2 &=& \frac{128 \pi  \alpha  \rho _0+192 \pi  \alpha  P\left(\rho _0\right)-3 \phi _0^2-2 \phi _0}{72 \alpha  (\phi _0+1) } \,,\nn \\
f_0 &=& 1\,, \quad f_2 = -\frac{128 \pi  \alpha  \rho _0+192 \pi  \alpha  P\left(\rho _0\right)+\phi _0 \left(3 \phi _0+4\right)}{72 \alpha  \left(\phi _0+1\right)} \,, \nn \\
\phi_2 &=& \frac{-16 \pi  \alpha  \rho _0+48 \pi  \alpha  P\left(\rho _0\right)+\phi _0}{36 \alpha } \,.
\eea
At the stellar surface, denoted by radius $R$, the pressure vanishes:
\be
p(R) = 0 \,.\label{surcond0}
\ee
To solve the vacuum field equations~(\ref{zeroordervac}), boundary conditions must also be specified both at the stellar surface and at spatial infinity. At the surface, continuity condition requires that the metric functions, scalar field, and their derivatives match smoothly across the interior and exterior regions:
\be
\lambda (R_\textup{in}) = \lambda (R_\textup{ext}) \,, \quad f (R_\textup{in}) = f (R_\textup{ext}) \,, \quad \phi (R_\textup{in}) = \phi (R_\textup{ext}) \,. \label{continue1} 
\ee
At asymptotic infinity, the leading falloff behavior of the ${\cal O}(\epsilon^0)$ functions $\lambda$, $f$, and $\phi$ is given by~\cite{Liu:2020yqa} 
\bea
\lim_{r \rightarrow \infty}  \lambda(r) &=& \ln \big(1 - \frac{2 M}{r} - \frac{\phi_c}{r} e^{-\frac{\mu}{\sqrt{3}}r} \big) \,, \nn \\
\lim_{r \rightarrow \infty}  f(r) &=& 1 - \frac{2 M}{r} +\phi_c \big(\frac{1}{r} + \frac{\mu}{\sqrt{3}} \big) e^{-\frac{\mu}{\sqrt{3}}r} \,, \nn \\
\lim_{r \rightarrow \infty}  \phi(r) &=&  \frac{\phi_c}{r} e^{-\frac{\mu}{\sqrt{3}}r} \,, \label{inftycond0}
\eea
where the Yukawa-type falloff originates from the weak-field limit of the theory, and $M$ and $\phi_c$ denote the mass and the scalar charge of the neutron star, respectively.

Having established the complete set of equilibrium equations~(\ref{zeroorder})--(\ref{zeroordervac}), along with the associated boundary conditions~(\ref{centercond1}), (\ref{surcond0}), and (\ref{inftycond0}), and continuity condition~(\ref{continue1}), we now proceed to the numerical integration of these equations. A shooting method is employed to integrate the interior equations~(\ref{zeroorder}) from the stellar center outward to the surface by using the initial conditions~(\ref{centercond1}). At the surface, the interior solution is smoothly matched to the exterior vacuum solution by imposing the continuity conditions specified in Eq.~(\ref{continue1}). The vacuum equations~(\ref{zeroordervac}) are then integrated outward from the stellar surface to a sufficiently large radius---typically ten times the stellar radius---where the numerical solution is compared with the asymptotic behavior~(\ref{inftycond0}). For a given central density $\rho_0$, the initial values of $\lambda_0$ and $\phi_0$ must be fine-tuned to ensure that the solution remains regular and free of divergences throughout both the stellar interior and the exterior vacuum region. The total order of the interior system~(\ref{zeroorder}) is five, while that of the vacuum system~(\ref{zeroordervac}) is four. However, in practice, the metric function $\lambda$ appears only through its derivative $\lambda'$ in the right-hand sides of the equations. (If the $g_{tt}$ component is expressed as $h = e^{\lambda}$, then both $h$ and $h'$ explicitly enter the equations.) Consequently, if one is only interested in computing macroscopic properties such as the stellar mass and radius, it is not necessary to solve for $\lambda$ itself, but only for its derivative $\lambda'$. In this sense, the effective total order of the interior and vacuum systems is effectively reduced to four and three, respectively. Finally, the mass $M$ and scalar charge $\phi_c$ of the neutron star can be extracted from the asymptotic falloffs of $f(r)$ and $\phi(r)$, respectively. In addition to using the shooting method to integrate the equilibrium equations from the stellar center to spatial infinity, alternative numerical approaches are also possible. For instance, one may integrate the equilibrium equations from the center outward to the stellar surface using the initial conditions specified by Eq.~(\ref{centercond1}), and separately integrate the vacuum field equations inward from a weak-field region using the initial conditions given in Eq.~(\ref{inftycond0}). By requiring the interior and exterior solutions to satisfy the continuity conditions at the stellar surface, as specified in Eq.~(\ref{continue1}) and boundary condition~(\ref{surcond0}), one can determine the values of $\lambda_0$, $\phi_0$, $M$, $R$, and $\phi_c$.

\subsection{Radial perturbation equations}

To study radial oscillations of neutron stars in Starobinsky gravity and its Gauss-Bonnet extension, we consider small time-dependent perturbations around a static equilibrium configuration. The ${\cal O}(\epsilon^0)$ components of the field equations determine the static background, while the ${\cal O}(\epsilon^1)$ components describe linear perturbations. We assume a harmonic time dependence for the radial displacement $\delta r$ and all associated perturbed quantities:
\begin{subequations} \label{timedependent}
\bea
\delta r(t, r) &=& \xi (r) e^{i \omega t} \,, \label{rtimedependent}\\
\delta \lambda(t, r) &=& \delta \lambda (r) e^{i \omega t} \,, \label{lambdatimedependent}\\
\delta f(t, r) &=& \delta f (r) e^{i \omega t} \,, \label{ftimedependent}\\
\delta \phi(t, r) &=& \delta \phi (r) e^{i \omega t} \,, \label{phitimedependent}\\
\delta p(t, r) &=& \delta p (r) e^{i \omega t} \,, \label{ptimedependent}\\
\delta \rho(t, r) &=& \delta \rho (r) e^{i \omega t} \,, \label{rhotimedependent}
\eea
\end{subequations}
where $\omega$ is the oscillation frequency.

In the case $\beta = 0$, corresponding to Starobinsky gravity, the ${\cal O}(\epsilon^1)$ components of the field equations yield the following system of linear perturbation equations:
\begin{subequations} \label{radialpert}
\bea
\delta E^t{}_t &\equiv&  \left(4 \alpha  \left(r f'+f-1\right)+r^2 \phi \right) \delta \phi +2 \alpha  r \left(r f'+4 f\right) \delta \phi ' +4 \alpha  f r^2 \delta \phi '' +32 \pi  \alpha  r^2 \delta \rho \nn \\
&\quad& +4 \alpha  \left(r^2 \phi ''+2 r \phi '+\phi +1\right) \delta f  +2 \alpha  r \left(r \phi '+2 \phi +2\right)\delta f'  =0 \,, \label{starobinskypert1}\\
\delta E^t{}_r &\equiv& 2 f r \delta \phi '- f r \lambda ' \delta \phi -16 \pi   r (P(\rho )+\rho )  \xi+\left(r \phi '+2 \phi +2\right)\delta f =0  \,, \label{starobinskypert2} \\
\delta E^r{}_r &\equiv& -32 \pi  \alpha r^2 P'(\rho ) \delta \rho + \left(\alpha  \left(4 \left(f r \lambda '+f-1\right)+4 e^{-\lambda } r^2 \omega ^2\right)+r^2 \phi \right) \delta \phi \nn \\
&\quad& +2 \alpha  f r \left(r \lambda '+4\right)\delta \phi '  +\alpha  \left(2 r \lambda ' \left(r \phi '+2\right)+4 \phi  \left(r \lambda '+1\right)+8 r \phi '+4\right) {\delta f} \nn \\
&\quad&  +2 \alpha  f r \left(r \phi '+2 \phi +2\right)\delta \lambda ' =0 \,, \label{starobinskypert3}\\
\delta E^\theta{}_\theta &\equiv& \left(\alpha  \left(f r \left(f' \left(r \lambda '+2\right)+f \left(2 \lambda '+2 r \lambda ''+r \lambda ' {}^2\right)\right)+4 f e^{-\lambda } r^2 \omega ^2\right)+f r^2 \phi \right)\delta \phi \nn \\ 
&\quad& +2 \alpha  f r  \left(r f'+f \left(r \lambda '+2\right)\right)\delta \phi ' +4 \alpha  f^2 r^2 \delta \phi '' +2 \alpha  f^2 r^2 (\phi +1) \delta \lambda '' \nn \\ 
&\quad& +\alpha  f r \left(r (\phi +1) f'+2 f \left(r \phi  \lambda '+r \lambda '+r \phi '+\phi +1\right)\right) \delta \lambda ' -32 \pi  \alpha  f r^2 P'(\rho ) \delta \rho \nn \\
&\quad& +\alpha  \big(f r \left(2 r (\phi +1) \lambda ''+2 \lambda ' \left(r \phi '+\phi +1\right)+r (\phi +1) \lambda ' {}^2+4 r \phi ''+4 \phi '\right) \nn \\
&\quad& -2 e^{-\lambda } r^2 \omega ^2 (\phi +1)\big) \delta f +\alpha  f r \left(\phi  \left(r \lambda '+2\right)+r \lambda '+2 r \phi '+2\right) \delta f' =0 \,, \label{starobinskypert4}\\
\delta E_\Phi &\equiv& \alpha\left(f \left(2 r^2 \lambda ''+r^2 \lambda ' {}^2+4 r \lambda '+4\right)-2 e^{-\lambda } r^2 \omega ^2\right) {\delta f} +\alpha  f r \left(r \lambda '+4\right) \delta f' \nn \\
&\quad& + f r^2 \delta \phi + 2 \alpha  f^2 r^2 \delta \lambda ''+\alpha  f r \left(r f'+2 f \left(r \lambda '+2\right)\right) \delta \lambda ' =0 \,. \label{starobinskypertphi}
\eea
\end{subequations}
From Eq.~(\ref{starobinskypertphi}), the Eulerian perturbation of the scalar field can be expressed as
\bea
\delta \phi &=& \alpha \left(\frac{2 e^{-\lambda } \omega ^2}{f}-\frac{2 r^2 \lambda ''+r^2 \lambda '{}^2+4 r \lambda '+4}{r^2}\right) \delta f -\frac{\alpha \left(r \lambda '+4\right)}{r} \delta f' \nn\\
&\quad& + \alpha  \left(-f'-\frac{2 f \left(r \lambda '+2\right)}{r}\right) \delta \lambda ' -2 \alpha  f \delta \lambda ''  \,.\label{starobinskypertphi2}
\eea
By substituting Eqs.~(\ref{starobinskyphi2}) and (\ref{starobinskypertphi2}) into the perturbation equations~(\ref{starobinskypert1})--(\ref{starobinskypert4}) and expanding all terms in powers of the coupling constant $\alpha$, one can explicitly verify that the equations reduce to those of GR in the limit $\alpha \to 0$.

For $\beta \ne 0$, corresponding to the Gauss-Bonnet extended Starobinsky gravity, the relevant perturbation expressions become more involved. In this general case, the radial perturbation equations arising from the ${\cal O}(\epsilon^1)$ components of the field equations form a set of linear ODEs, which can be written as:
\bea
\xi^{\prime} &=& A_{11}  \xi +A_{12}  \delta \lambda^{\prime} + A_{13} \delta f + A_{14} \delta \phi   \,, \nn\\
\delta \lambda^{\prime\prime} &=& A_{21}  \xi +A_{22}  \delta \lambda^{\prime} + A_{23}  \delta f + A_{24}  \delta \phi \,, \nn\\
\delta f^{\prime} &=& A_{31}  \xi +A_{32}  \delta \lambda^{\prime} + A_{33}  \delta f + A_{34} \delta \phi \,, \nn\\
\delta \phi^{\prime} &=& A_{41}  \xi +A_{42} \delta \lambda^{\prime} + A_{43}  \delta f + A_{44}  \delta \phi \,.   \label{1storder1}
\eea 
Here, the coefficient functions $A_{ij}$ depend on the radial coordinate $r$, the background functions ($\lambda, f, \phi, \rho$) and their derivatives, as well as the squared frequency $\omega^2$. These coefficients are introduced to make the system more compact and to highlight its structure. Their explicit expressions are provided in the Supplemental Material for reference.

The corresponding vacuum perturbation equations are obtained by setting $\tilde{\rho} = 0$ and $\tilde{p} = 0$ in the ${\cal O}(\epsilon^1)$ components of the field equations. In the case of $\beta = 0$, corresponding to Starobinsky gravity, this procedure reduces Eq.~(\ref{radialpert}) accordingly. As a result, the perturbation equations simplify significantly in the vacuum region. The original set of linear ODEs describing radial oscillations in the stellar interior(\ref{1storder1}) reduces, in the vacuum, to a system of two first-order linear ODEs:
\bea
\delta f^{\prime} &=& \tilde{A}_{11}  \delta f + \tilde{A}_{12} \delta \phi \,, \nn\\
\delta \phi^{\prime} &=& \tilde{A}_{21}  \delta f + \tilde{A}_{22}  \delta \phi \,.  \label{1stordervac}
\eea
The expressions for the vacuum coefficients $\tilde{A}_{ij}$ are also provided in the Supplemental Material.

\subsection{Boundary conditions inside the stars}

To solve the perturbation equations~(\ref{1storder1})--(\ref{1stordervac}), we also need derive the corresponding boundary conditions. Near the stellar center, by performing power series expansions for the perturbation functions ($\xi, \delta \lambda, \delta f, \delta \phi, \delta\rho$), the regular behavior of the solutions can be expressed as follows:
\bea
\lim_{r \rightarrow 0}  \xi(r) &=& \xi_0 + \xi_1 r +{\cal O}(r^2) \,, \nn \\
\lim_{r \rightarrow 0} \delta \lambda(r) &=& \delta \lambda_0 +\delta \lambda_1 r + \delta \lambda_2 r^2 +{\cal O}(r^3) \,, \nn \\
\lim_{r \rightarrow 0} \delta f (r) &=& \delta f_0 + \delta f_1 r+ \delta f_2 r^2 +{\cal O}(r^3)\,, \nn \\ 
\lim_{r \rightarrow 0} \delta\phi(r) &=& \delta\phi_0 + \delta\phi_1 r  +{\cal O}(r^2)\,, \nn \\
\lim_{r \rightarrow 0} \delta\rho(r) &=& \delta\rho_0 +{\cal O}(r^1)\,. \label{centercond2}
\eea
In this expansion, the nonzero coefficients include three free parameters $(\xi_1, \delta \lambda_0, \delta \phi_0)$, along with $(\delta \lambda_2, \delta f_2, \delta \rho_0)$. The leading-order density perturbation is given by
\be
\delta \rho_0 = -3 \xi _1 (P(\rho _0)+\rho _0) \,, \label{deltalambdaf}
\ee
which coincides with the result in GR. The general expressions for $\delta \lambda_2$ and $\delta f_2$ are summarized in the Supplemental Materials. In the case of $\beta = 0$, corresponding to Starobinsky gravity, they reduce to
\bea
\delta\lambda_2 &=& -\frac{ \left(24 \alpha  e^{-\lambda _0} \omega ^2 \left(\phi _0+1\right)+128 \pi  \alpha  \rho _0+192 \pi  \alpha  P\left(\rho _0\right)+3 \phi _0^2+6 \phi _0+2\right)}{72 \alpha  \left(\phi _0+1\right){}^2} \delta \phi _0 \nn \\
&\quad& -\frac{8 \pi \left(P\left(\rho _0\right)+\rho _0\right) \left(3 P'\left(\rho _0\right)+2\right)}{3 \left(\phi _0+1\right)}  \xi _1 \,, \\
\delta f_2 &=& \frac{\left(24 \alpha  e^{-\lambda _0} \omega ^2 \left(\phi _0+1\right)+128 \pi  \alpha  \rho _0+192 \pi  \alpha  P\left(\rho _0\right)-3 \phi _0^2-6 \phi _0-4\right)}{72 \alpha  \left(\phi _0+1\right){}^2} \delta \phi _0  \nn  \\
&\quad& +\frac{8 \pi  \left(P\left(\rho _0\right)+\rho _0\right) \left(3 P'\left(\rho _0\right)+2\right)}{3 \left(\phi _0+1\right)}  \xi _1 \,.
\eea
To be physically acceptable solution, the Lagrangian perturbation of the pressure $p$ at the stellar surface in the perturbed configuration must vanish, i.e.,
\be
\Delta p (R) = 0 \,. \label{pertboundary}
\ee
where $\Delta$ denotes the Lagrangian perturbation, which differs from the Eulerian perturbation $\delta$ used earlier.
The Eulerian perturbation $\delta p$ refers to the change in pressure at a fixed spatial point, while the Lagrangian perturbation $\Delta p$ corresponds to changes of pressure $p$ measured by an observer moving with the fluid, i.e., an observer who would be located at radial coordinate $r$ in the unperturbed configuration but at $r + \delta r$ in the perturbed configuration. The two are related via~\cite{Misner:1973prb}
\be
\Delta p (t, r) =  p (t, r +\delta r (t, r)) - p(r) = \delta p (t, r) + p^\prime(r) \delta r (t, r) \,. \label{Lagrangiandef}
\ee
By substituting the time-dependent forms of $ \delta p (t, r) $ and $ \delta r (t, r) $ from Eqs.~(\ref{rtimedependent}) and (\ref{ptimedependent}), along with the EOS relations~(\ref{eos0})--(\ref{eos1}) into Eq.~(\ref{Lagrangiandef}), the boundary condition~(\ref{pertboundary}) becomes
\be
\Delta p (t, R) = e^{i\omega t} P'(\rho)\left(\delta \rho (R) + \rho^\prime(R) \xi (R) \right) \Longrightarrow  \delta \rho (R) + \rho^\prime(R) \xi (R) = 0 \,. \label{surcond1}
\ee
This condition is standard in the analysis of radial oscillations in Newtonian gravity and GR~\cite{Misner:1973prb,Shapiro:1983du}, and remains valid in modified gravity theories. However, due to modifications in the gravitational field equations, the specific expressions for $\delta \rho$, $\rho'$, and $\xi$ are altered accordingly. For instance, in the case $\beta = 0$ (Starobinsky gravity), the Lagrangian pressure perturbation takes the form
\bea
\Delta p &=& \frac{2  (P(\rho )+\rho )}{r} \xi + \left( \frac{3 f \phi +3 f+8 \pi  r^2 P(\rho )+\phi +1}{8 \pi  r^2 \lambda '+32 \pi  r}-\frac{r \phi ^2}{64 \pi  \alpha  \left(r \lambda '+4\right)} \right) \delta \lambda ' \nn\\
&\quad& + \left(\frac{8 \pi  r^2 P(\rho )-(3 f-1) (\phi +1)}{16 \pi  f r^2}-\frac{\phi ^2}{128 \pi  \alpha  f}\right) \delta f \nn\\
&\quad& + \left(\frac{\phi }{32 \pi  \alpha }-\frac{-f \left(r^2 \left(\lambda '\right)^2+8 r \lambda '+4\right)-4 e^{-\lambda } r^2 \omega ^2+4}{32 \pi  r^2}\right) \delta \phi \,. \label{1storder2}
\eea
By substituting Eqs.~(\ref{starobinskyphi2}) and (\ref{starobinskypertphi2})  into Eq.~(\ref{1storder2}) and expanding in powers of the small parameter $\alpha$, one can explicitly verify that the equations reduce to those of GR in the limit $\alpha \to 0$. For $\beta \ne 0$, corresponding to the Gauss-Bonnet extension of Starobinsky gravity, the expression for the Lagrangian pressure perturbation becomes considerably more involved and is therefore provided in the Supplemental Material.

\subsection{Boundary conditions outside the stars}

To solve the vacuum perturbation equations~(\ref{1stordervac}), it is also necessary to derive the corresponding boundary conditions. By imposing continuity at the stellar surface, we require that the perturbation functions match smoothly across the interior and exterior regions:
\be
\delta f (R_\textup{in}) = \delta f  (R_\textup{ext}) \,, \quad \delta \phi (R_\textup{in}) = \delta \phi (R_\textup{ext}) \,.  \label{continue2}
\ee
At spatial infinity, the asymptotic behavior of the perturbation functions is given by
\be
\lim_{r \rightarrow \infty}  \delta f (r) = \lim_{r \rightarrow \infty}  \delta \phi (r) = 0 \,.\label{inftycond1}
\ee

Having established the complete set of perturbation equations~(\ref{1storder1})--(\ref{1stordervac}), together with the associated boundary conditions~(\ref{deltalambdaf})--(\ref{inftycond1}), we now turn to the numerical integration of these perturbation equations. 
The equilibrium equations can first be solved using either of the two methods discussed earlier, followed by a shooting procedure for the perturbation equations. Alternatively, one may simultaneously solve the equilibrium and perturbation equations using a unified shooting scheme. Both approaches are equivalent.

In our implementation, we first solve the equilibrium equations~(\ref{zeroorder})--(\ref{zeroordervac}) independently to determine appropriate initial values for $\lambda_0$ and $\phi_0$, as well as to obtain physical quantities such as the mass $M$, radius $R$, and scalar hair $\phi_c$. Once the background solution is obtained, we proceed to simultaneously integrate both the equilibrium~(\ref{zeroorder})--(\ref{zeroordervac}) and perturbation~(\ref{1storder1})--(\ref{1stordervac}) systems. The shooting method is employed to integrate the interior equations outward from the stellar center to the surface. When Eqs.~(\ref{surcond0}) and ~(\ref{pertboundary}) are satisfied---i.e., when the integration reaches the stellar surface---the interior solutions for both the background and perturbation functions are smoothly matched to their vacuum counterparts by imposing the continuity conditions specified in Eqs.~(\ref{continue1}) and (\ref{continue2}). The vacuum equations---both equilibrium~(\ref{zeroordervac}) and perturbation~(\ref{1stordervac})---are then integrated outward from the stellar surface to a sufficiently large radius, where the perturbation functions asymptotically vanish by imposing Eq.~(\ref{inftycond1}).

For a given central density $\rho_0$, since the appropriate initial values for $\lambda_0$ and $\phi_0$ have already been determined, it remains necessary to determinate the values of $\delta\lambda_0$, $\xi_1$, $\delta\phi_0$, and $\omega^2$ to ensure that the perturbation solutions remain regular and free of divergences throughout both the stellar interior and the exterior vacuum region. 
It is worth noting that the value of $\delta \lambda_0$ does not affect the values of other parameters or the perturbation solutions. This is because $\delta \lambda$ does not explicitly appear in the perturbation equations---only its derivative, $\delta \lambda'$, does. As a result, although the full system described by Eq.~(\ref{1storder1}) is formally fifth order, the effective order of the system is reduced to four. This feature arises from our choice of expressing the $g_{tt}$ component of the metric as $g_{tt} = e^{\tilde{\lambda}}$. Moreover, as seen from Eq.~(\ref{deltalambdaf}), the choice of the free parameter $\xi_1$ only needs to ensure that the density perturbation $\delta \rho_0$ remains finite, which is consistent with the situation in GR~\cite{Misner:1973prb}. The value of $\xi_1$ merely determines the amplitude of the radial oscillation and does not affect the resulting frequency. This point is also confirmed by our numerical calculations presented later in the paper. Therefore, for simplicity, we set $\delta \lambda_0 = \xi_1 = 1$ in this work. With these choices, the shooting procedure for integrating the perturbation equations effectively reduces to a two-dimensional parameter search over $(\delta \phi_0, \omega^2)$. Once an appropriate range for these parameters is identified through preliminary trial-and-error explorations, the desired numerical accuracy can always be achieved.

In addition, we nondimensionalize the equations during the numerical computation. Throughout this work, we use the subscript “$\star$” to denote the characteristic dimensions of various physical quantities. These dimensions scale as
\be
M_\odot \sim r_\star \sim \phi_{\star} \sim p_\star^{-\frac12} \sim \rho_\star^{-\frac12} \sim \alpha_\star^{\frac12} \sim \mu_\star^{-1} \sim \beta_\star^{\frac12} \sim \omega_\star^{-1}  \,, \label{dimension}
\ee
with the solar mass $M_\odot$ chosen as the reference unit. This scaling allows us to recast Eqs.~(\ref{zeroorder})--(\ref{zeroordervac}) and (\ref{1storder1})--(\ref{1stordervac})  into a dimensionless form suitable for numerical treatment.
The characteristic dimensional units, normalized by $M_\odot$ and expressed in centimeter-gram-second (CGS) units, are given by:
\bea
r_\star &=& \frac{G M_{\odot}}{c^2} = 1.48 \times 10^{5}\text{cm} \,, \nn \\
\phi_{\star} &=& M_{\odot} = 1.99 \times 10^{33}\text{g} \,, \nn \\
\rho_\star &=& \dfrac{c^6}{G^3 M_{\odot}^2} = 6.18 \times 10^{17}\text{g}\cdot\text{cm}^{-3} \,,  \nn \\ 
p_\star &=& \dfrac{c^8}{G^3 M_{\odot}^2} = 5.55 \times 10^{38}\text{g}\cdot\text{cm}^{-1}\cdot\text{s}^{-2} \,,  \nn \\
\omega_\star &=& \dfrac{c^3}{G M_{\odot}} = 2.03 \times 10^{5}\text{s}^{-1} \,.
\eea

\section{Results for the stellar structure and radial stability}  \label{Stability}

Having established all the equations to be solved and the corresponding boundary conditions, and having introduced the numerical methods employed---specifically, the shooting method---we now specify the EOS used for neutron stars. In this section, we adopt a realistic baryonic EOS, SLy (Skyrme Lyon)~\cite{Douchin:2001sv}, whose analytical form is provided in Ref.~\cite{Haensel:2004nu}. In addition, a specific polytropic EOS is employed in Appendix~\ref{appendix} to demonstrate the consistency of radial oscillation behavior between the Jordan frame and the Einstein frame of the equivalent scalar-tensor gravity corresponding to Starobinsky gravity. The main goal of this work is to investigate the radial stability of neutron stars within Starobinsky gravity, and to extend this analysis to its Gauss-Bonnet modification, thereby exploring a broader class of modified gravities. To this end, we consider six representative sets of coupling constants based on their deviation from GR: (1) $\alpha = 0$, $\beta = 0$; (2) $\alpha = 10 \alpha_\star $, $\beta = 0$; (3) $\alpha = 100 \alpha_\star$, $\beta = 0$; (4) $\alpha = 1000 \alpha_\star$, $\beta = 0$; (5) $\alpha = 1000 \alpha_\star$, $\beta = -10\beta$; and (6) $\alpha = 1000 \alpha_\star$, $\beta = 50 \beta_\star$. Set (1) corresponds to GR, while sets (2)--(4) represent Starobinsky gravity with increasing coupling strength. Sets (5) and (6) correspond to the Gauss-Bonnet extension. In all cases, the Schwarzschild black hole remains a valid solution in the exterior vacuum. 

\begin{figure}[]
\includegraphics[width=0.95\linewidth]{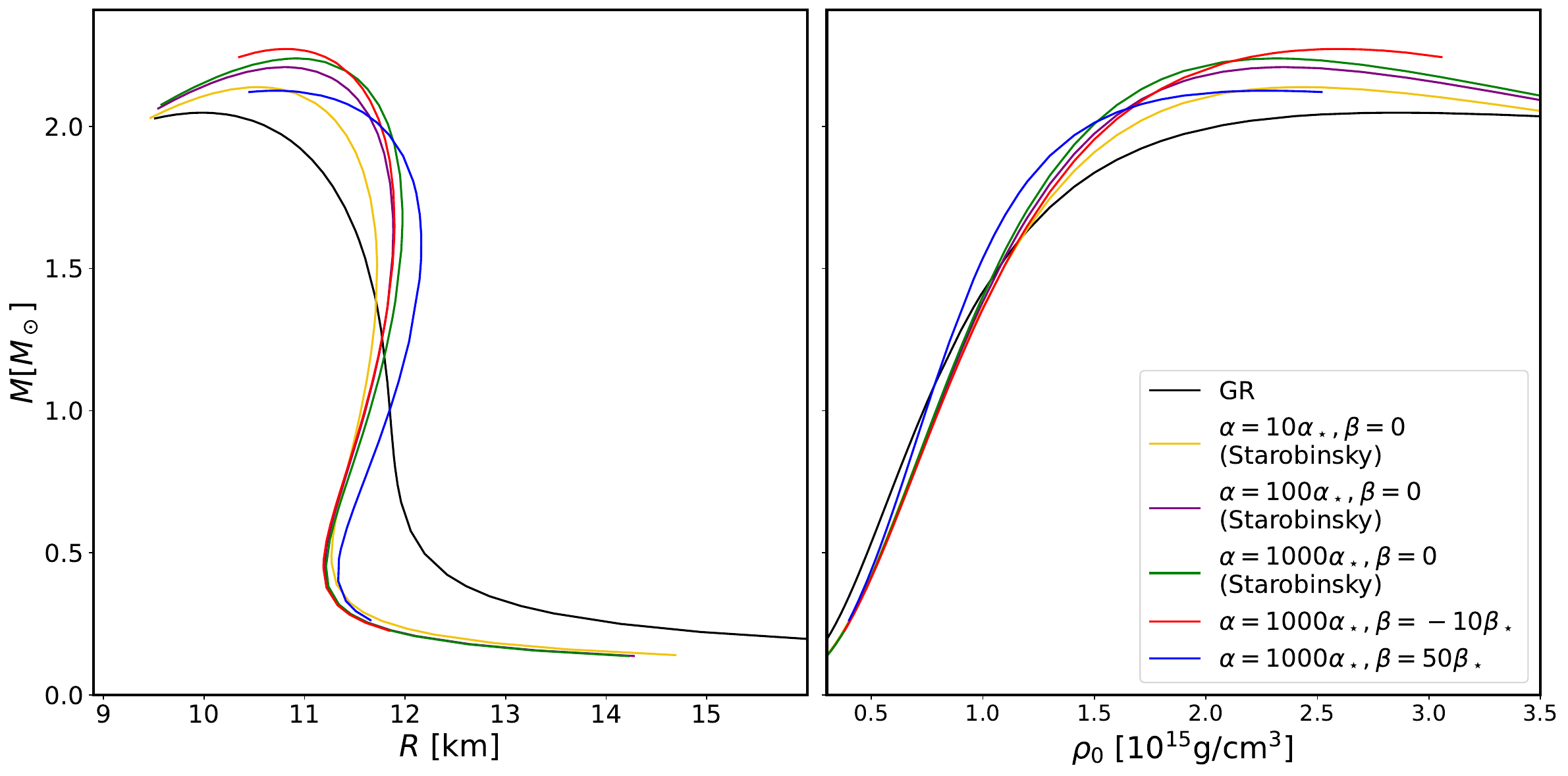}
\caption{\label{fig:RM}The $M-R$ and $M-\rho_0$ relations in various gravitational frameworks, obtained using the SLy EOS. Different colors correspond to different values of the coupling constants $\alpha$ and $\beta$.}
\end{figure}

\begin{figure}[]
\includegraphics[width=0.75\linewidth]{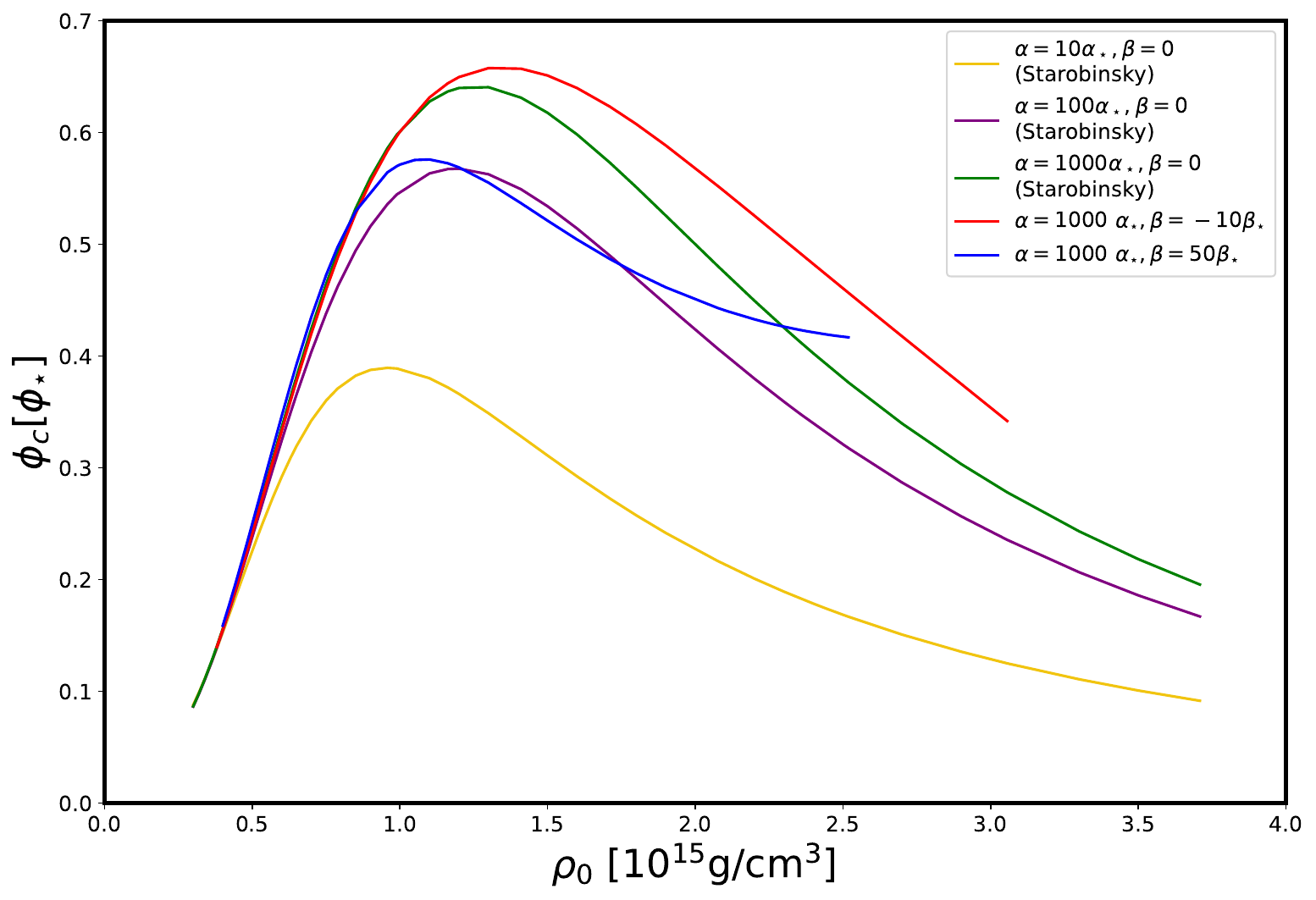}
\caption{\label{fig:phicrho0}The relationship between the scalar charge and the central density ($\phi_c-\rho_0$) in various gravitational frameworks, with the SLy EOS. Different colors represent different coupling constants, $\alpha$ and $\beta$.}
\end{figure}

We compute the equilibrium structures and dynamical behavior of neutron stars across a range of central densities within the six gravitational frameworks. The mass-radius ($M-R$) and mass-central density ($M-\rho_0$) relations for these models are shown in Fig.~\ref{fig:RM}.
As shown in Fig.~\ref{fig:RM}, the inclusion of high-curvature terms can significantly affect the mass and radius of stellar models. A more detailed analysis of these effects can be found in Ref.~\cite{Liu:2024wvw}. To further illustrate the scalarization behavior, we plot in Fig.~\ref{fig:phicrho0} 
the relation between the scalar charge $\phi_c$ and the central density $\rho_0$ for these models. As shown in Fig.~\ref{fig:phicrho0}, the scalar charge $\phi_c$ typically exhibits a peak within the range of central densities. This behavior is analogous to the existence of a maximum mass at a characteristic central density.

Having obtained the equilibrium configurations, we now proceed to analyze the radial oscillation behavior of the stellar models. It is worth noting that in the study of radial oscillations in neutron stars, the stability of a given configuration can be assessed by examining the sign of the squared frequency $\omega^2$ associated with the fundamental mode. As indicated in Eq.~(\ref{timedependent}), a positive value of $\omega^2$ corresponds to an oscillatory solution, implying dynamical stability, whereas a negative value leads to an exponentially growing mode, indicating instability.

In GR, the radial oscillation equation inside a star reduces to a second-order ODE~\cite{Misner:1973prb}. Due to Birkhoff's theorem, there are no dynamical perturbation equations in the vacuum region, and the exterior spacetime remains described by the Schwarzschild metric. The interior radial oscillation equation can be cast into the form of a standard Sturm-Liouville problem. Subject to the appropriate boundary conditions at the stellar center and surface~(\ref{pertboundary})---where perturbation functions such as $\delta f, \delta \lambda', \delta {\cal R}$ naturally decay to zero before reaching the surface---the system admits an infinite discrete spectrum of eigenvalues, corresponding to an infinite set of squared frequencies $\omega^2$. The lowest $\omega^2$ corresponds to the fundamental mode, while the higher represent overtone modes.

In Starobinsky gravity and its Gauss-Bonnet extension, the situation is markedly different. Due to the higher-derivative nature of higher-curvature gravity, Birkhoff's theorem is no longer valid. Consequently, the vacuum region outside the star can support nontrivial dynamical perturbations, governed by the second-order equation~(\ref{1stordervac}). Moreover, the vacuum solution is no longer necessarily described by the Schwarzschild metric. These higher-curvature modifications alter the mathematical structure of the perturbation problem, rendering it qualitatively distinct from the standard Sturm-Liouville formulation encountered in GR.

In our actual computations, we employ the shooting method to solve the radial perturbation equations~(\ref{1storder1})--(\ref{1stordervac}), along with the equilibrium equations~(\ref{zeroorder})--(\ref{zeroordervac}), in both the interior and exterior regions of the star. With appropriate choices of the free parameter $\delta\phi_0$ and squared frequencies $\omega^2$, the numerical integration can be stabilized, and the boundary conditions at the stellar center and surface~(\ref{pertboundary}) can be satisfied. Similar to the case in GR, one can obtain multiple sets of squared frequencies, ordered from smallest to largest, which correspond to the fundamental mode $\omega_0^2$ as well as higher overtones $\omega_n^2$ ($n = 1, 2, \dots$).
However, once the continuity conditions~(\ref{continue2}) at the stellar surface are imposed and the surface values of the perturbation functions are used as initial conditions for integrating the vacuum equations outward to spatial infinity under the requirement of asymptotic flatness~(\ref{inftycond1}), it turns out that only the fundamental mode $\omega_0^2$ yields perturbation functions $\delta f$ and $\delta \phi$ that exponential decay to zero at infinity. The overtone modes fail to satisfy the boundary conditions at spatial infinity, as their corresponding perturbation functions do not asymptotically vanish but instead exhibit oscillatory behavior.

\begin{figure}[]
\includegraphics[width=0.95\linewidth]{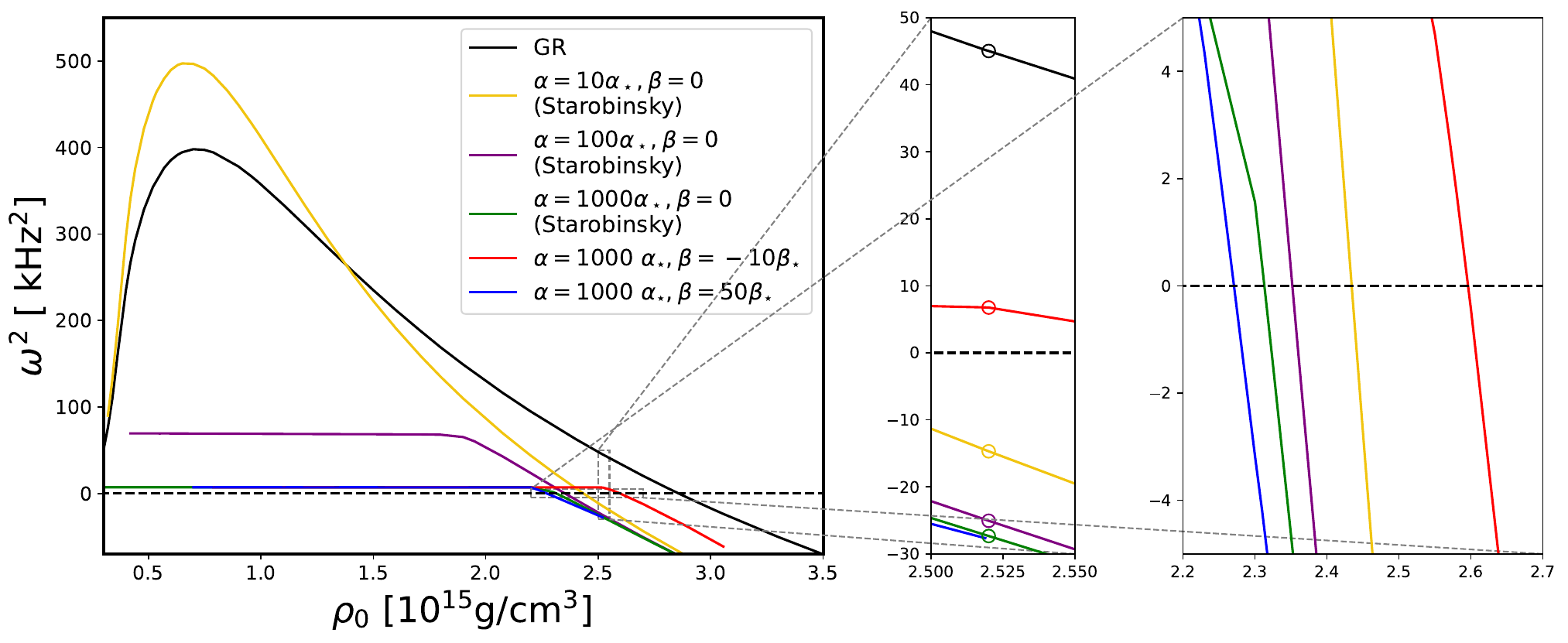}
\caption{\label{fig:wsqrho}The relationship between frequency-squared for the fundamental oscillation mode and central density ($\omega^2-\rho_0$) in various gravitational frameworks, with the SLy EOS. Different colors represent different coupling constants, $\alpha$ and $\beta$.}
\end{figure}

We plot the relationship between the frequency-squared $\omega^2$ for the fundamental oscillation mode and central density $\rho_0$ for different coupling constants in Fig.~\ref{fig:wsqrho}.
From Fig.~\ref{fig:wsqrho}, we observe that modifications to GR, which induce deviations in the exterior spacetime from the Schwarzschild geometry, alter the stability of stellar models. For instance, as shown in the middle panel of Fig.~\ref{fig:wsqrho}, a stellar model with a central density of $\rho_0 = 4.08 \times 10^{-3} \rho_{\star}$ ($2.52 \times 10^{15}  \textup{g/cm}^3$) is stable within the GR framework. However, when the ${\cal R}^2$ modification is introduced (i.e., $\alpha \ne 0$, $\beta = 0$), and the exterior spacetime deviates from the Schwarzschild geometry, the same model may gradually become unstable. This instability strengthens with increasing $\alpha$, corresponding to a larger departure from GR. Interestingly, when the nature of the modification changes---such as by incorporating the Gauss-Bonnet extension of the ${\cal R}^2$ term (with $\alpha \ne 0$ and $\beta \ne 0$)---the model may regain stability, even though the overall deviation from GR is more significant (e.g., $\beta < 0$, which amplifies the effect of the ${\cal R}^2$ correction).

\begin{figure}[]
\includegraphics[width=0.95\linewidth]{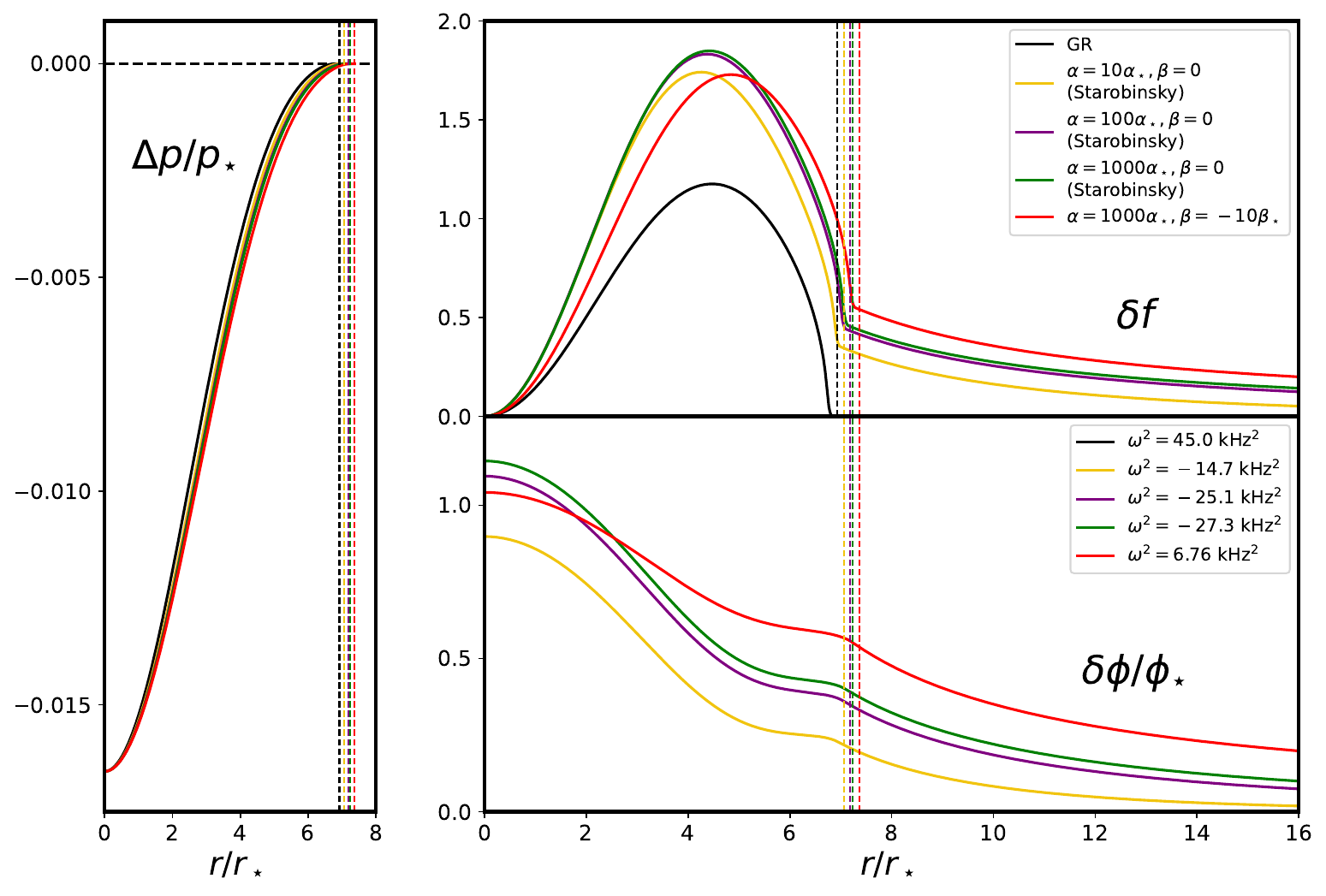}
\caption{\label{fig:oscillationsol}Numerical solutions for $\delta f(r)$, $\delta \phi(r)$, and $\Delta p(r)$ of a stellar model with the SLy EOS and central density $\rho_0 = 4.08 \times 10^{-3} \rho_{\star}$ $(2.52 \times 10^{15} \textup{g/cm}^3)$ in various gravitational frameworks. Different colors correspond to different coupling constants $\alpha$ and $\beta$. The vertical dashed line represents the stellar radius.}
\end{figure}

To verify that the computed frequency indeed corresponds to the fundamental mode, we present in Fig.~\ref{fig:oscillationsol} the perturbation profiles of $\delta f$, $\delta \phi$, and Lagrangian perturbation of pressure $\Delta p$ for the stellar model with a central density of $\rho_0 = 2.52 \times 10^{15}  \textup{g/cm}^3$ shown in the middle panel of Fig.~\ref{fig:wsqrho}.
The functions $\delta f$ and $\delta \phi$ are obtained by solving Eqs.~(\ref{1storder1})--(\ref{1stordervac}), while $\Delta p$ is evaluated by substituting these solutions into its explicit expression, Eq.~(\ref{surcond1}) (or Eq.~(\ref{1storder2}) in the case of $\beta = 0$).
From Fig.~\ref{fig:oscillationsol}, we observe that all perturbation functions $(\Delta p, \delta f, \delta \phi)$ are node-free. The corresponding values of the free parameters used in the shooting method for the stellar model $\rho_0 = 4.08 \times 10^{-3} \rho_{\star}$ ($2.52 \times 10^{15}  \textup{g/cm}^3$) shown in Fig.~\ref{fig:oscillationsol} are summarized in Table~\ref{table1}.
\begin{table}[htbp]
\centering
\begin{tabular}{cccccccc}
\hline
$\alpha [\alpha_\star]$  & $\beta [\beta_\star]$  & $\rho_0 [10^{-3} \rho_\star]$      & $\lambda_0$  &  $\phi_0 [\phi_\star]$ & $\xi_1 [r_\star]$  & $\delta\phi_0 [\phi_\star]$  &  $\omega^2  [10^{-3}\omega_\star^2]$   \\
\hline
$0 $                & 0                  & $4.08  $    & -2.30     & null      &     1     & null          & $1.093$   \\
$10 $               & 0                  & $4.08 $    & -2.35     & 0.0345  &     1     & 0.90         & $-0.356$   \\
$10 $               & 0                  & $4.08 $    & -2.35     & 0.0345  &     2     & 1.80         & $-0.356$   \\
$10 $               & 0                  & $4.08 $    & -2.35     & 0.0345  &     10    & 8.98         & $-0.356$   \\
$10^2 $            & 0                  & $4.08 $    & -2.37     & 0.0712  &     1     & 1.10         & $-0.608$   \\
$10^3 $            & 0                  & $4.08 $    & -2.39     & 0.0854  &     1     & 1.15         & $-0.663$   \\
$10^3 $            & -10                & $4.08 $    & -2.39     & 0.168  &     1     & 1.04         & $0.164$   \\
$10^2 $            & 0                  & $1.94 $    & -1.22     & 0.159  &     1     & 1.93         & $1.674 $   \\
$10^2 $            & 0                  & $2.43 $    & -1.56     & 0.158  &     1     & 1.45         & $1.663$   \\
$10^2 $            & 0                  & $2.91 $    & -1.85     & 0.140  &     1     & 1.09         & $1.650$   \\
$10^3 $             & 0                  & $2.91 $    & -1.86     & 0.161  &     1     & 1.64         & $0.171$   \\
$10^3 $             & -10                & $2.91 $    & -1.85     & 0.208  &     1     & 2.12         & $0.171$   \\
$10^3 $             & 50                 & $2.91 $    & -1.83     & 0.0515  &     1     & 0.92         & $0.170$   \\
\hline
\end{tabular}
\caption{Free parameters $\lambda_0$, $\phi_0$, $\xi_1$, $\delta \phi_0$, and the squared fundamental frequency $\omega^2$ for different central densities $\rho_0$ and coupling constants $\alpha$ and $\beta$. }\label{table1}
\end{table}
For the stellar model with $\alpha = 10$, we varied the value of $\xi_1$ (e.g., $\xi_1 = 2$ and $10$), and confirmed that the resulting frequencies remain unchanged, consistent with our previous statement. The corresponding values of the parameter $\delta \phi_0$ for each case are summarized in Table~\ref{table1}.
Moreover, in gravitational frameworks where Birkhoff's theorem is violated, the exterior spacetime dynamically responds to radial oscillations of the stellar fluid. In contrast, under GR, the exterior geometry remains static during such oscillations. For the same stellar model, greater deviations from the Schwarzschild geometry result in more significant variations in the exterior spacetime induced by fluid oscillations.

\begin{figure}[]
\includegraphics[width=0.9\linewidth]{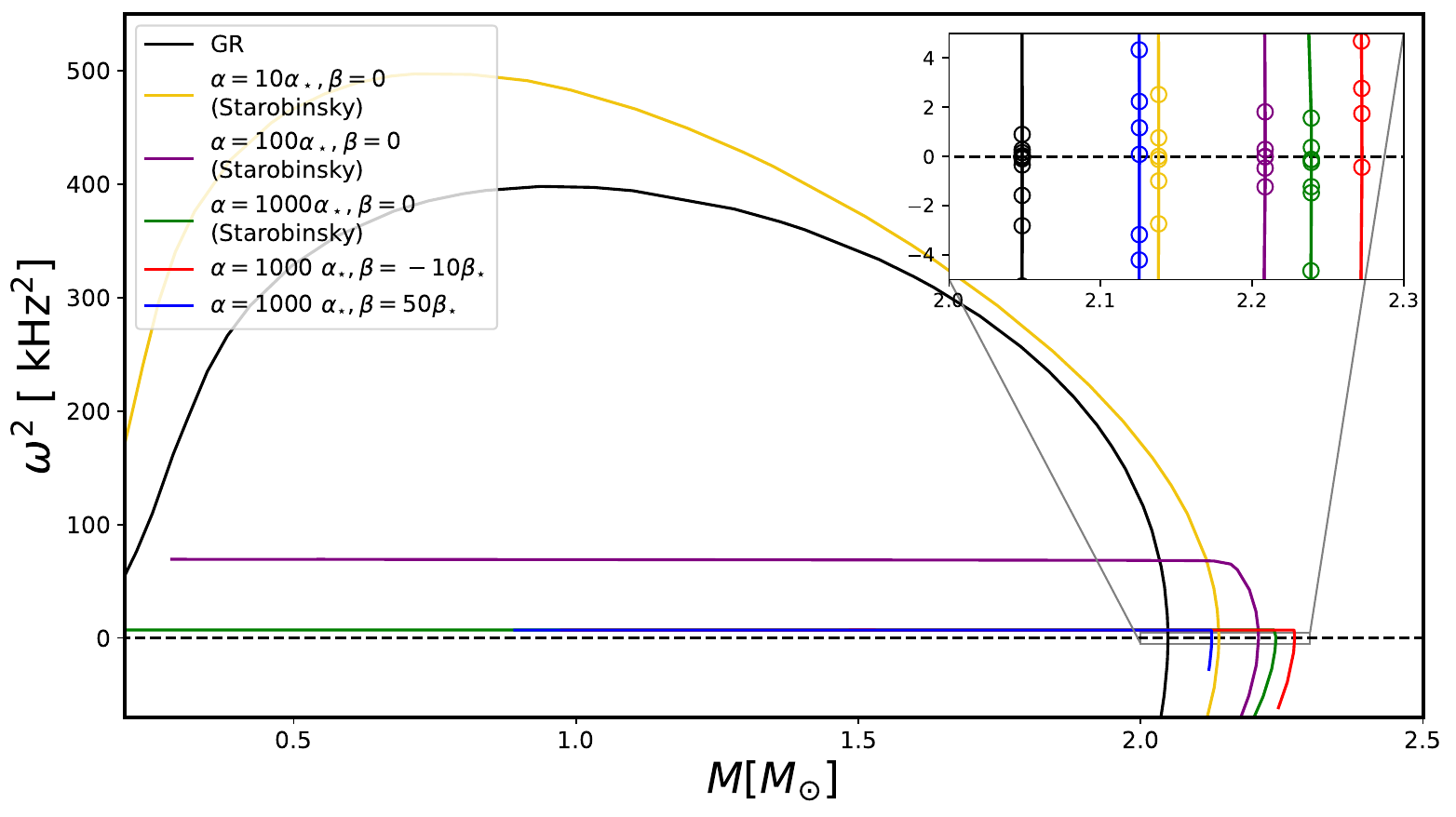}
\caption{\label{fig:wMrelation}The relationship between frequency-squared for the fundamental oscillation mode and mass ($\omega^2-M$) in various gravitational frameworks, with the SLy EOS. Different colors represent different coupling constants, $\alpha$ and $\beta$. }
\end{figure}

Next, we investigate the impact of gravitational modifications on the global properties of stellar models within a certain range of central densities. To this end, we further plot the relationship between the frequency-squared $\omega^2$ for the fundamental oscillation mode and mass $M$ for different coupling constants in Fig.~\ref{fig:wMrelation}. 
We observe a new feature in stellar models with relatively low central densities when the coupling constant $\alpha$ is large---specifically, in cases where the exterior spacetime significantly deviates from the Schwarzschild geometry, such as $\alpha = 100\alpha_\star$ or $1000\alpha_\star$, the squared frequency $\omega^2$ of the fundamental oscillation mode becomes nearly independent of central density $\rho_0$, as shown in Figs.~\ref{fig:wsqrho} and \ref{fig:wMrelation}.
Moreover, when $\alpha$ is sufficiently large (e.g., $\alpha = 1000\alpha_\star$), the influence of a nonzero $\beta$ on the oscillation frequency also becomes negligible. To facilitate independent verification of this behavior, we provide three representative stellar models with increasing central densities  $ 1.94 \times 10^{-3} \rho_{\star}$ $(1.20 \times 10^{15} \textup{g/cm}^3)$, $2.43 \times 10^{-3} \rho_{\star}$ $(1.50 \times 10^{15} \textup{g/cm}^3)$, $2.91 \times 10^{-3} \rho_{\star}$ $(1.80 \times 10^{15} \textup{g/cm}^3)$  for the case $\alpha = 100\alpha_\star$, along with their corresponding initial values for the shooting method in Table~\ref{table1}. For the case $\alpha = 1000 \alpha_\star$, the initial values of the free parameters for a representative stellar model with central density $ 2.91 \times 10^{-3} \rho_{\star}$ $(1.80 \times 10^{15} \textup{g/cm}^3)$ are also provided in Table~\ref{table1} for validation purposes, with different values of the Gauss-Bonnet coupling constant: $\beta = 0$, $-10 \beta_\star$, and $50 \beta_\star$.

On the other hand, some global properties of stellar models with high central densities remain unchanged. In particular, for neutron stars composed of baryonic EOS, the transition from stability to instability in radial oscillations typically occurs at the maximum-mass configuration in GR~\cite{Haensel:2007yy, Ferrari:2020nzo}. To verify whether this feature persists in Starobinsky gravity and its Gauss-Bonnet extension, we increased the density of data points near the central density corresponding to the maximum mass. We found that the static stability criterion $dM/d\rho_0 > 0$ continues to hold up to the second decimal place in mass for all values of the coupling constants $\alpha$ and $\beta$ considered. 
To better illustrate this feature as shown in Fig.~\ref{fig:wMrelation}, we summarize in Table~\ref{table2} the squared frequencies, masses, and central densities of stellar models in the vicinity of the maximum-mass configuration.
\begin{table}[htbp]
	\centering
	\begin{tabular}{ccccc}
		\hline
		$\alpha [\alpha_\star]$  & $\quad\beta [\beta_\star]\quad$  & $\rho_0\,[10^{15} \text{g}/\text{cm}^{-3}]$    & $\quad M\,[M_{\odot}]\quad$   &  $\omega^2$[kHz${}^2$]   \\
		\hline
		$0 $                & 0                  &   $2.70  $       & 2.047           & $20.17$   \\
		$0 $                & 0                  &   $2.80 $       & 2.048           & $7.184$   \\
		$0 $                & 0                  &   $2.90 $       & 2.048           & $-5.260$   \\
		$0 $               & 0                  &   $3.00 $       & 2.047           & $-17.20$   \\
		$10 $               & 0                  &   $2.30 $       & 2.136           & $24.37$   \\
		$10 $               & 0                  &   $2.40 $       & 2.138           & $6.063$   \\
		$10 $              & 0                  &   $2.45 $       & 2.138           & $-2.736$   \\
		$10 $              & 0                  &   $2.52 $       & 2.137           & $-14.69$   \\
		$10^2 $              & 0                  &   $2.30 $       & 2.208           & $7.934$   \\
		$10^2 $              & 0                  &   $2.35 $       & 2.209           & $0.287$   \\
		$10^2 $              & 0                  &   $2.36 $       & 2.209           & $-1.233$   \\
		$10^2 $              & 0                  &   $2.50 $       & 2.205           & $-22.13$   \\
		$10^3 $              & 0                  &   $2.20 $       & 2.236           & $7.069$   \\
		$10^3 $              & 0                  &   $2.30 $       & 2.239           & $1.561$   \\
		$10^3 $              & 0                  &   $2.35 $       & 2.239           & $-4.644$   \\
		$10^3 $              & 0                  &   $2.40 $       & 2.238           & $-11.19$   \\
		$10^3 $             & -10                  &   $2.50 $       & 2.271           & $6.975$   \\
		$10^3 $             & -10                  &   $2.55 $       & 2.272           & $4.686$   \\
		$10^3 $             & -10                &   $2.60 $       & 2.272           & $-0.436$   \\
		$10^3 $             & -10                &   $2.70 $       & 2.271           & $-12.37$   \\
		$10^3 $             & 50                &   $2.20 $       & 2.125           & $6.969$   \\
		$10^3 $             & 50                 &   $2.25 $       & 2.126           & $2.232$   \\
		$10^3 $             & 50                 &   $2.30 $       & 2.126           & $-3.176$   \\
		$10^3 $             & 50                 &   $2.35 $       & 2.125           & $-8.684$   \\
		\hline
	\end{tabular}
\caption{Squared frequencies $\omega^2$, masses $M$ and central densities $\rho_0$ of neutron star models near the maximum-mass configuration, for different coupling constants $\alpha$ and $\beta$. }\label{table2}
\end{table}

\section{Summary and discussion} \label{Conclusion}

In this work, we investigate the radial oscillations of neutron stars in Starobinsky gravity, using the baryonic SLy EOS to model the stellar fluid. The main analysis is performed in the Jordan frame of the scalar-tensor gravity equivalent to Starobinsky gravity. In Appendix~\ref{appendix}, we verify the consistency between the Jordan and Einstein frames by employing a specific class of polytropic EOS and demonstrating that both frames yield identical results. Furthermore, we extend our analysis to a class of Gauss-Bonnet modifications of Starobinsky gravity.

Our main findings are as follows: First, gravitational modifications can alter the stability of individual stellar models. Second, the higher-derivative nature of Starobinsky gravity and its Gauss-Bonnet extension invalidates Birkhoff’s theorem, thereby allowing the exterior spacetime to respond dynamically to the radial oscillations of the stellar fluid. Third, for a range of stellar models with relatively low central densities, the frequency $\omega$ of the fundamental mode becomes approximately independent of the central density $\rho_0$ when the curvature-squared corrections are significant, i.e., when the coupling constant $\alpha$ is large. Fourth, the static stability criterion $dM/d\rho_0 > 0$ from GR remains approximately valid in both Starobinsky gravity and its Gauss-Bonnet extension.

While our analysis focuses on Starobinsky gravity and its Gauss-Bonnet extension, these models are merely specific examples within the broader class of higher-derivative gravity theories. Whether our findings can be generalized to other modified gravity frameworks remains an open question. In particular, it is unclear whether the observed decoupling of the fundamental mode frequency from the central density for low-density stellar models persists in other theories, and whether the static stability criterion continues to hold for models with high central densities. Further investigation along these lines may shed light on the nature of gravity and reveal potential novel characteristics of modified gravity theories in strong-field astrophysical systems.

\appendix 

\section{Radial oscillations in Starobinsky gravity within the Einstein frame} \label{appendix}

In this appendix, we study the radial oscillations of neutron stars in the Einstein frame of the scalar–tensor theory equivalent to Starobinsky gravity, as a comparison to the analysis performed in the Jordan frame in the main text.

The action of Starobinsky gravity in the Jordan frame is given by Eq.~(\ref{action}), where the corresponding Lagrangian takes the form of Eq.~(\ref{edgb}) with $\beta = 0$.The Einstein frame is obtained by introducing a new scalar field $\hat{\Phi}$ and a conformally related metric $\hat{g}_{\mu\nu}$ via
\bea
\hat{\Phi} &=& \frac{\sqrt{3}}{2} \ln (\Phi +1) \,, \\
\hat{g}_{\mu\nu} &=& (\Phi +1) {g}_{\mu\nu} \,. \label{transformmetric}
\eea
In terms of these variables, the action in the Einstein frame becomes
\be
S = \frac{1}{16 \pi } \int{ d^4 x \sqrt{-\hat{g}} \hat{L}} +S_\textup{m}(e^{-\frac{2}{\sqrt{3}}\hat{\Phi}} \hat{g}_{\mu\nu}, \chi)  \,, \label{action}
\ee
with the gravitational Lagrangian
\be
\hat{L} =\hat{\cal R} - 2 (\nabla \hat{\Phi} )^2 - V(\hat{\Phi}) \,,\quad \textup{with}\quad V(\hat{\Phi}) = \frac{1}{4\alpha} \left(1 -  e^{-\frac{2}{\sqrt{3}}\hat{\Phi}} \right)^2 \,.
\ee
The EOMs  in the Einstein frame are given by
\bea
\hat{E}_{\mu\nu} &\equiv& \hat{G}_{\mu\nu} - 2 \left[\nabla_\mu \hat{\Phi}\nabla_\nu \hat{\Phi} -\frac12\hat{g}_{\mu\nu}(\nabla \hat{\Phi} )^2  \right] + \frac{1}{8\alpha} \left(1 -  e^{-\frac{2}{\sqrt{3}}\hat{\Phi}} \right)^2 \hat{g}_{\mu\nu} = 8\pi \hat{T}_{\mu\nu} \,,\\
\hat{E}_{\hat{\Phi}} &\equiv& 4\ \Box \hat{\Phi} -\frac{1}{\sqrt{3}\alpha} \left(1 -  e^{-\frac{2}{\sqrt{3}}\hat{\Phi}} \right) e^{-\frac{2}{\sqrt{3}}\hat{\Phi}} = \frac{16\pi }{\sqrt{3}} \hat{T} \,.
\eea
The energy-momentum tensor $\hat{T}_{\mu\nu}$ and its trace $\hat{T}$ in the Einstein frame are related to their Jordan-frame counterparts by
\bea
\hat{T}_{\mu\nu} &=& -\frac{2}{\sqrt{-\hat{g}}} \frac{\partial S_\textup{m}}{\partial \hat{g}^{\mu\nu}} = (\Phi +1)^{-2} \left(-\frac{2}{\sqrt{-{g}}} \frac{\partial S_\textup{m}}{\partial {g}^{\mu\nu}} \right) \frac{\partial {g}^{\mu\nu}}{\partial \hat{g}^{\mu\nu}} = (\Phi +1)^{-1} T_{\mu\nu} \,, \\
\hat{T} &=& \hat{T}_{\mu\nu} \hat{g}^{\mu\nu} = (\Phi +1)^{-2} T \,.
\eea
For a perfect fluid, the density, pressure, and four-velocity transform between the two frames as
\be
\hat{\rho} = (\Phi +1)^{-2} \rho \,,\quad \hat{p} = (\Phi +1)^{-2} p \,,\quad \hat{u}^t = (\Phi +1)^{-1/2} {u}^t \,.
\ee

After presenting the transformation relations between the gravitational field, scalar field, and matter quantities (such as density and pressure) in the two frames, we omit the hat notation for Einstein-frame quantities in what follows to avoid notational clutter. However, since the EOS is defined in the Jordan frame, we continue to express the density $\rho$ and pressure $p$ in terms of their Jordan-frame values throughout the analysis. This convention is widely adopted in the literature, such as Ref.~\cite{Yazadjiev:2014cza}.

We are now ready to investigate the radial oscillations of neutron stars in Starobinsky gravity within the Einstein frame, following the same perturbative approach outlined in Sec.~\ref{framework}. The ${\cal O}(\epsilon^0)$ components of the field equations for $E^\mu{}_\nu$ and $E_\Phi$ in the Einstein frame reduce to the following background system:
\begin{subequations} \label{TOVein}
\bea
E^t{}_t &\equiv& -\frac{e^{-\frac{4 \phi }{\sqrt{3}}} \left(e^{\frac{2 \phi }{\sqrt{3}}}-1\right)^2}{8 \alpha }-\frac{r f'+f-1}{r^2}-f \left(\phi '\right)^2-8 \pi  \rho  e^{-\frac{4 \phi }{\sqrt{3}}} =0 \,, \label{starobinsky1e}\\
E^r{}_r &\equiv& \frac{e^{-\frac{4 \phi }{\sqrt{3}}} \left(e^{\frac{2 \phi }{\sqrt{3}}}-1\right)^2}{8 \alpha }+\frac{f-1}{r^2}+\frac{f \lambda '}{r}-f \left(\phi '\right)^2-8 \pi  p e^{-\frac{4 \phi }{\sqrt{3}}} =0 \,,\label{starobinsky2e}\\
E^\theta{}_\theta &\equiv& \frac{1}{2} r \left(f'+f r \lambda ''\right)+\frac{1}{4} r \lambda ' \left(r f'+2 f\right)+\frac{1}{4} f r^2 \left(\lambda '\right)^2+f r^2 \left(\phi '\right)^2 \nn\\ 
&\quad& -8 \pi  p r^2 e^{-\frac{4 \phi }{\sqrt{3}}}+\frac{r^2 e^{-\frac{4 \phi }{\sqrt{3}}} \left(e^{\frac{2 \phi }{\sqrt{3}}}-1\right)^2}{8 \alpha } =0 \,,\label{starobinsky3e}\\
E_\Phi &\equiv& \phi ' \left(\frac{f'}{2 f}+\frac{\lambda '}{2}+\frac{2}{r}\right)-\frac{e^{-\frac{4 \phi }{\sqrt{3}}} \left(e^{\frac{2 \phi }{\sqrt{3}}}-1\right)}{4 \sqrt{3} \alpha  f} +\frac{4 \pi  e^{-\frac{4 \phi }{\sqrt{3}}}}{\sqrt{3} f} (\rho-3 p) +\phi '' =0 \,. \label{starobinskyphie}
\eea
\end{subequations} 
These equations are consistent with those derived in Ref.~\cite{Yazadjiev:2014cza}. The corresponding vacuum field equations are obtained by setting $\rho = 0$ and $p = 0$ in Eq.~(\ref{TOVein}).
To study the radial stability of neutron stars in the Einstein frame, we now consider the ${\cal O}(\epsilon^1)$ perturbation equations obtained by linearizing the field equations with respect to the perturbed metric and scalar field. The resulting system governs the behavior of the perturbation functions $\delta f$, $\delta \lambda$, and $\delta \phi$, along with the fluid displacement function $\xi$.
The ${\cal O}(\epsilon^1)$ components of $\delta E^\mu{}_\nu$ and $\delta E_\Phi$ take the following form:
\begin{subequations} \label{radialpertein}
\bea
\delta E^t{}_t &\equiv& 48 \alpha  f r^2 e^{\frac{4 \phi }{\sqrt{3}}} \phi ' \delta \phi ' 
+24 \alpha e^{\frac{4 \phi }{\sqrt{3}}} \left(r^2 \left(\phi '\right)^2+1\right) \delta f
+192 \pi  \alpha  r^2 \delta \rho
 \nn \\
&\quad& + \left(4 \sqrt{3} r^2 \left(e^{\frac{2 \phi }{\sqrt{3}}}-1\right)-256 \sqrt{3} \pi  \alpha  \rho  r^2\right) \delta \phi 
+24 \alpha  r e^{\frac{4 \phi }{\sqrt{3}}} \delta f'  =0 \,, \label{starobinskypert1e}\\
\delta E^t{}_r &\equiv& \delta f+2  f r \phi '  \delta \phi -8 \pi  r e^{-\frac{4 \phi }{\sqrt{3}}} (P(\rho )+\rho )  \xi  =0  \,, \label{starobinskypert2e} \\
\delta E^r{}_r &\equiv& 48 \alpha  f^2 r^2 e^{\frac{4 \phi }{\sqrt{3}}} \phi ' \delta \phi ' 
-24 \alpha  f^2 r e^{\frac{4 \phi }{\sqrt{3}}} \delta \lambda ' +192 \pi  \alpha  f r^2 P'(\rho ) \delta \rho \nn \\
&\quad& + \left(3 r^2 \left(e^{\frac{2 \phi }{\sqrt{3}}}-1\right)^2-24 \alpha  \left(8 \pi  r^2 P(\rho )+e^{\frac{4 \phi }{\sqrt{3}}}\right)\right) \delta f  \nn \\
&\quad& + \left(-256 \sqrt{3} \pi  \alpha  f r^2 P(\rho )-4 \sqrt{3} f r^2 \left(e^{\frac{2 \phi }{\sqrt{3}}}-1\right)\right) \delta \phi   =0 \,, \label{starobinskypert3e}\\
\delta E^\theta{}_\theta &\equiv& \left[\alpha  \left(3 f \left(r^2 \left(\phi '\right)^2+1\right)+24 \pi  r^2 e^{-\frac{4 \phi }{\sqrt{3}}} P(\rho )+3\right)-\frac{3}{8} r^2 e^{-\frac{4 \phi }{\sqrt{3}}} \left(e^{\frac{2 \phi }{\sqrt{3}}}-1\right)^2\right] \delta f' \nn \\ 
&\quad& + \Big[3 \alpha  f \left(f \left(r^2 \left(\phi '\right)^2-1\right)+16 \pi  r^2 e^{-\frac{4 \phi }{\sqrt{3}}} P(\rho )-8 \pi  \rho  r^2 e^{-\frac{4 \phi }{\sqrt{3}}}+3\right) \nn \\
&\quad& -\frac{9}{8} f r^2 e^{-\frac{4 \phi }{\sqrt{3}}} \left(e^{\frac{2 \phi }{\sqrt{3}}}-1\right)^2\Big] \delta \lambda ' +6 \alpha  f^2 r \delta \lambda ''  -96 \pi  \alpha   f r e^{-\frac{4 \phi }{\sqrt{3}}} P'(\rho ) \delta \rho \nn \\ 
&\quad& +\left[128 \pi  \sqrt{3} \alpha  f r e^{-\frac{4 \phi }{\sqrt{3}}} P(\rho )+2 \sqrt{3} f r e^{-\frac{4 \phi }{\sqrt{3}}} \left(e^{\frac{2 \phi }{\sqrt{3}}}-1\right)\right]\delta \phi +24 \alpha  f^2 r  \phi ' \delta \phi ' \nn \\
&\quad& +\Bigg[3 \alpha  e^{-\frac{8 \phi }{\sqrt{3}}} \Bigg(8 \pi  r^2 P(\rho ) \left(f e^{\frac{4 \phi }{\sqrt{3}}} \left(r^2 \left(\phi '\right)^2+5\right)+8 \pi  \rho  r^2-e^{\frac{4 \phi }{\sqrt{3}}}\right) \nn \\
&\quad& +e^{\frac{4 \phi }{\sqrt{3}}} \left(f r^2 \left(\phi '\right)^2+f+1\right) \left(f e^{\frac{4 \phi }{\sqrt{3}}} \left(r^2 \left(\phi '\right)^2+1\right)+8 \pi  \rho  r^2-e^{\frac{4 \phi }{\sqrt{3}}}\right)\Bigg)\Big/({f r} ) \nn \\
&\quad& -{3 r e^{-\frac{8 \phi }{\sqrt{3}}} \left(e^{\frac{2 \phi }{\sqrt{3}}}-1\right)^2 \left(2 f e^{\frac{4 \phi }{\sqrt{3}}}-4 \pi  r^2 P(\rho )+4 \pi  \rho  r^2-e^{\frac{4 \phi }{\sqrt{3}}}\right)}\Big/({4 f}) \nn \\
&\quad& -{3 r^3 e^{-\frac{8 \phi }{\sqrt{3}}} \left(e^{\frac{2 \phi }{\sqrt{3}}}-1\right)^4}\Big/({64 \alpha  f})-6 \alpha  r \omega ^2 e^{-\lambda}\Bigg] \delta f =0 \,, \label{starobinskypert4e}\\
\delta E_\Phi &\equiv&  \Bigg[\frac{e^{-\frac{4 \phi }{\sqrt{3}}} \left(3 e^{\frac{4 \phi }{\sqrt{3}}} \phi ' \left(f r^2 \left(\phi '\right)^2+f-1\right)+24 \pi  \sqrt{3} r P(\rho )-8 \pi  \rho  r \left(\sqrt{3}-3 r \phi '\right)\right)}{f r} \nn \\
&\quad& +\frac{e^{-\frac{4 \phi }{\sqrt{3}}} \left(e^{\frac{2 \phi }{\sqrt{3}}}-1\right) \left(3 r \left(e^{\frac{2 \phi }{\sqrt{3}}}-1\right) \phi '+4 \sqrt{3}\right)}{8 \alpha  f}\Bigg] \delta f + 3 \phi ' \delta f' +6 f \delta \phi '' \nn \\
&\quad& + \left[\frac{6 e^{-\frac{4 \phi }{\sqrt{3}}}  \left(f e^{\frac{4 \phi }{\sqrt{3}}}+4 \pi  r^2 P(\rho )-4 \pi  \rho  r^2+e^{\frac{4 \phi }{\sqrt{3}}}\right)}{r}  -\frac{3 r e^{-\frac{4 \phi }{\sqrt{3}}} \left(e^{\frac{2 \phi }{\sqrt{3}}}-1\right)^2 }{4 \alpha } \right] \delta \phi '  \nn \\
&\quad& + \left[\frac{e^{-\frac{4 \phi }{\sqrt{3}}} \left(e^{\frac{2 \phi }{\sqrt{3}}}-2\right)}{\alpha }+6 e^{-\lambda } \omega ^2+32 \pi  e^{-\frac{4 \phi }{\sqrt{3}}} (3 P(\rho )-\rho )\right] \delta \phi  \nn \\
&\quad& -8 \sqrt{3} \pi  e^{-\frac{4 \phi }{\sqrt{3}}} \left(3 P'(\rho )-1\right)  \delta \rho  +3 f \phi ' \delta \lambda ' =0 \,. \label{starobinskypertphie}
\eea
\end{subequations}
This system describes the interior region of the neutron star. The corresponding equations for the vacuum exterior are obtained by setting $\delta \rho = 0$ and $\delta p = 0$ in Eq.~(\ref{radialpertein}).

To solve the equilibrium and perturbation equations~(\ref{TOVein})--(\ref{radialpertein}), it is necessary to derive the corresponding boundary conditions. The regular behavior of the background fields ($\lambda$, $f$, $\phi$, $\rho$) and perturbation variables ($\xi$, $\delta\lambda$, $\delta f$, $\delta\phi$, $\delta\rho$) near the stellar center can be obtained through power-series expansions, as shown in Eqs.~(\ref{centercond1}) and(\ref{centercond2}). The independent free parameters characterizing the expansions at the center are ($\rho_0$, $\lambda_0$, $\phi_0$, $\xi_1$, $\delta\lambda_0$, $\delta\phi_0$). The leading nonvanishing expansion coefficients take the following forms:
\bea
\lambda_2 &=& -\frac{e^{-\frac{4 \phi _0}{\sqrt{3}}} \left(-32 \pi  \alpha  \rho _0-96 \pi  \alpha  P\left(\rho _0\right)-2 e^{\frac{2 \phi _0}{\sqrt{3}}}+e^{\frac{4 \phi _0}{\sqrt{3}}}+1\right)}{24 \alpha } \,,\\
f_0 &=& 1 \,, \quad  f_2 = -\frac{e^{-\frac{4 \phi _0}{\sqrt{3}}} \left(64 \pi  \alpha  \rho _0-2 e^{\frac{2 \phi _0}{\sqrt{3}}}+e^{\frac{4 \phi _0}{\sqrt{3}}}+1\right)}{24 \alpha }  \,,\\
\phi_2 &=& \frac{e^{-\frac{4 \phi _0}{\sqrt{3}}} \left(-16 \pi  \alpha  \rho _0+48 \pi  \alpha  P\left(\rho _0\right)+e^{\frac{2 \phi _0}{\sqrt{3}}}-1\right)}{24 \sqrt{3} \alpha } \,,\\
 \delta f_2 &=& 8 \pi  e^{-\frac{4 \phi _0}{\sqrt{3}}} \left(P\left(\rho _0\right)+\rho _0\right) \xi _1 -\frac{ e^{-\frac{4 \phi _0}{\sqrt{3}}} \left(-16 \pi  \alpha  \rho _0+48 \pi  \alpha  P\left(\rho _0\right)+e^{\frac{2 \phi _0}{\sqrt{3}}}-1\right)}{6 \sqrt{3} \alpha } \delta \phi _0 \,,\\
\delta \phi_2 &=& \Bigg[-\frac{1}{6} e^{-\lambda _0} \omega ^2  +\frac{2}{9} \pi  e^{-\frac{4 \phi _0}{\sqrt{3}}} \left(9 \rho _0 P'\left(\rho _0\right)+3 P\left(\rho _0\right) \left(3 P'\left(\rho _0\right)-5\right)+\rho _0\right) \nn \\
&\quad& -\frac{e^{-\frac{4 \phi _0}{\sqrt{3}}} \left(e^{\frac{2 \phi _0}{\sqrt{3}}}-2\right)}{36 \alpha } \Bigg]  \delta \phi _0  -\frac{2 \pi  e^{-\frac{4 \phi _0}{\sqrt{3}}} \left(P\left(\rho _0\right)+\rho _0\right) \left(3 P'\left(\rho _0\right)-1\right)}{\sqrt{3}}  \xi _1 \,,\\
\delta \lambda_2  &=& \left(\frac{4 \pi  e^{-\frac{4 \phi _0}{\sqrt{3}}} \left(9 P\left(\rho _0\right) \left(P'\left(\rho _0\right)-1\right)+\rho _0 \left(9 P'\left(\rho _0\right)-1\right)\right)}{3 \sqrt{3}}-\frac{e^{-\frac{4 \phi _0}{\sqrt{3}}} \left(e^{\frac{2 \phi _0}{\sqrt{3}}}-1\right)}{6 \sqrt{3} \alpha }\right) \delta \phi _0 \nn \\
&\quad& -4 \pi  e^{-\frac{4 \phi _0}{\sqrt{3}}} \left(P\left(\rho _0\right)+\rho _0\right) \left(3 P'\left(\rho _0\right)+1\right)  \xi _1 \,,\\
\delta\rho_0&=&  \left(P\left(\rho _0\right)+\rho _0\right) \left(\sqrt{3} \delta \phi _0-3 \xi _1\right) \,.
\eea 
In addition to these regularity conditions at the center, one must impose the surface boundary conditions~(\ref{surcond0}) and~(\ref{pertboundary}). Specifically, the Lagrangian perturbation of pressure $\Delta p$ at the surface is given by:
\bea
\Delta p &=&  \left(\frac{e^{\frac{4 \phi }{\sqrt{3}}} \left(f \left(-3 r^2 \left(\phi '\right)^2+2 \sqrt{3} r \phi '+3\right)+3\right)+24 \pi  r^2 P(\rho )}{48 \pi  f r^2}-\frac{\left(e^{\frac{2 \phi }{\sqrt{3}}}-1\right)^2}{128 \pi  \alpha  f}\right)\delta f \nn \\
&\quad&  + \Bigg[\frac{3 \phi ' \left((f-1) e^{\frac{4 \phi }{\sqrt{3}}}-8 \pi  r^2 P(\rho )\right)-3 f r^2 e^{\frac{4 \phi }{\sqrt{3}}} \left(\phi '\right)^3+2 \sqrt{3} f r e^{\frac{4 \phi }{\sqrt{3}}} \left(\phi '\right)^2+32 \pi  \sqrt{3} r P(\rho )}{24 \pi  r} \nn \\
&\quad& +\frac{\left(e^{\frac{2 \phi }{\sqrt{3}}}-1\right) \left(3 r \left(e^{\frac{2 \phi }{\sqrt{3}}}-1\right) \phi '+4 \sqrt{3}\right)}{192 \pi  \alpha }\Bigg]\delta \phi   -\frac{f e^{\frac{4 \phi }{\sqrt{3}}} \phi '}{4 \pi }  \delta \phi ' +\frac{f e^{\frac{4 \phi }{\sqrt{3}}}}{8 \pi  r} \delta \lambda ' \,.
\eea
At the stellar surface, the continuity conditions~(\ref{continue1}) and~(\ref{continue2}) must also be enforced.
It is worth noting that the radius defined in the Jordan frame, $R_\textup{J}$, is related to that in the Einstein frame, $R_\textup{E}$, through the conformal factor evaluated at the stellar surface:
\be
R_\textup{J} =  e^{-\frac{1}{\sqrt{3}} \hat{\Phi}(R_\textup{E})} R_\textup{E}  \,.
\ee
This relation arises from the transformation~(\ref{transformmetric}), which indicates that the radial coordinate in the two frames differs by a  conformal factor. Additionally, asymptotic conditions at spatial infinity are imposed, as specified in~(\ref{inftycond0}) and~(\ref{inftycond1}). Due to the conformal transformation between two frames, the asymptotic asymptotic behavior (\ref{inftycond0}) can be simply expressed as~\cite{Yazadjiev:2014cza}
\be
\lim_{r \rightarrow \infty}  \lambda(r) = \ln \big(1 - \frac{2 M}{r} \big) \,, \quad
\lim_{r \rightarrow \infty}  f(r) = 1 - \frac{2 M}{r}  \,, \quad
\lim_{r \rightarrow \infty}  \phi(r) = 0 \,. \label{inftycond0e}
\ee
To verify the consistency of the squared frequency $\omega^2$ for the fundamental mode between the two frames, and to facilitate comparison with GR, we employ a class of polytropic EOS as described in Table A.18 of Ref.~\cite{Kokkotas:2000up}:
\be
p = \kappa \rho^{1+\frac{1}{n}} \,,\quad \textup{with} \quad \kappa =45.86 \kappa_\star \,,\quad n = 1 \,, \label{polytropic}
\ee
where $\kappa_\star$ sets the dimensional scale of $\kappa$, and in CGS units is given by
\be
\kappa_\star = c^{2-6/n} G^{3/n} M_{\odot}^{2/n} \,.
\ee
We have computed the masses, radii, and fundamental-mode frequencies of the stellar models discussed in the literature using our numerical code in three settings: GR, the Jordan frame, and the Einstein frame of Starobinsky gravity with $\alpha = 10$. Our results are summarized in Table~\ref{table3}, and they confirm that the squared frequencies obtained in the Jordan and Einstein frames are consistent with each other.\begin{table}[htbp]
\centering
\begin{tabular}{cccccccc}
\hline
$\rho_c$ [$10^{15}$ g/cm$^3$] 
& $M_{\text{GR}} [M_{\odot}] $    & $R_{\text{GR}} $ [km] & ${\omega_{\text{GR}}}/({2\pi})$  [kHz]
& $M [M_{\odot}] $                       & $R_{\text{J}}$ [km]      & $R_{\text{E}}$ [km]               
& ${\omega}/({2\pi})$ [kHz]  \\
\hline
		$5.700 $ & 1.351 & 7.517 & -0.339${}^\dagger$        & 1.440 & 7.746 & 7.914 & -5.437${}^\dagger$       \\
		$5.650 $ & 1.351 & 7.535 & 0.166         & 1.440 & 7.767 & 7.936 & -4.722${}^\dagger$      \\
		$5.600 $ & 1.351 & 7.553 & 0.351         & 1.441 & 7.788 & 7.958 & -4.001${}^\dagger$      \\
		$5.500 $ & 1.351 & 7.590 & 0.561         & 1.441 & 7.831 & 8.003 & -2.563${}^\dagger$      \\
		$5.300 $ & 1.350 & 7.666 & 0.835         & 1.442 & 7.918 & 8.095 & 0.248        \\
		$5.000 $ & 1.348 & 7.787 & 1.125         & 1.440 & 8.053 & 8.237 & 0.963        \\
		$4.000 $ & 1.326 & 8.256 & 1.750         & 1.405 & 8.540 & 8.744 & 1.923        \\
		$3.000 $ & 1.266 & 8.862 & 2.138         & 1.301 & 9.090 & 9.299 & 2.464        \\
		$2.000 $ & 1.126 & 9.673 & 2.320         & 1.084 & 9.702 & 9.892 & 2.718        \\
		$1.500 $ & 0.998 & 10.19 & 2.299         & 0.912 & 10.03 & 10.20 & 2.691        \\
		$1.000 $ & 0.802 & 10.81 & 2.147         & 0.683 & 10.37 & 10.50 & 2.493        \\
\hline
\end{tabular}
\caption{Comparison of the mass $M$, radius $R$, and fundamental-mode frequency $\omega$ of selected stellar models with the polytropic EOS~(\ref{polytropic}), computed in GR and in the Jordan and Einstein frames of Starobinsky gravity with $\alpha = 10$. Entries marked with $\dagger$ denote squared frequency values, $\omega^2/(2\pi)^2$. GR results are cross-checked with Table A.18 of Ref.~\cite{Kokkotas:2000up}, where $\nu_0 = \omega_{\text{GR}}/(2\pi)$.}
\label{table3}
\end{table}

\begin{acknowledgments}
We are grateful to Hong-Bo Li, and Rui Xu for useful discussions. We are particularly indebted to the referee for their many constructive and valuable comments.
S.L. and H.Y. were supported in part by the National Natural Science Foundation of China (No. 12105098, No. 12481540179, No. 12075084, No. 11690034, No. 11947216, and No. 12005059) and the Natural Science Foundation of Hunan Province (No. 2022JJ40264), and the innovative research group of Hunan Province under Grant No. 2024JJ1006.

\end{acknowledgments}

% Create the reference section using BibTeX:
%\bibliography{basename of .bib file}

\end{document}